# Unravelling CO adsorption on model single-atom catalysts

**Authors:** Jan Hulva,[1]† Matthias Meier,[1,2]† Roland Bliem,[1]‡ Zdenek Jakub,[1] Florian Kraushofer,[1] Michael Schmid,[1] Ulrike Diebold,[1] Cesare Franchini,[2,3] Gareth S. Parkinson[1]*

**Affiliations:**

[1]Institute of Applied Physics, TU Wien, Vienna, Austria.

[2]Computational Materials Physics, University of Vienna, Vienna, Austria.

[3]Alma Mater Studiorum – Universita di Bologna, Bologna, Italy.

*Correspondence to: parkinson@iap.tuwien.ac.at

†These authors contributed equally to this work

‡Current address: Advanced Research Center for Nanolithography, Science Park 106, 1098XG Amsterdam, The Netherlands

**Abstract:** Understanding how the local environment of a "single-atom" catalyst affects stability and reactivity remains a significant challenge. We present an in-depth study of $Cu_1$, $Ag_1$, $Au_1$, $Ni_1$, $Pd_1$, $Pt_1$, $Rh_1$, and $Ir_1$ species on $Fe_3O_4(001)$; a model support where all metals occupy the same 2-fold coordinated adsorption site upon deposition at room temperature. Surface science techniques revealed that CO adsorption strength at single metal sites differs from the respective metal surfaces and supported clusters. Charge transfer into the support modifies the d-states of the metal atom and the strength of the metal-CO bond. These effects could strengthen the bond (as for $Ag_1$-CO) or weaken it (as for $Ni_1$-CO), but CO-induced structural distortions reduce adsorption energies from those expected based on electronic structure alone. The extent of the relaxations depends on the local geometry and could be predicted by analogy to coordination chemistry.

**One Sentence Summary:** A systematic study shows that single-atom catalysts differ from supported metal clusters, and have properties defined by the local coordination environment.

**Main Text:** The reactivity of oxide-supported metal nanoparticle catalysts is traditionally understood using the d-band model, which was developed for extended metal surfaces (*1*). Such a picture cannot be applied to so-called "single-atom" catalysts (*2-8*), because the isolated metal atoms are stabilized by chemical bonds to the support and are often charged. In this regard, single-atom catalysis systems resemble coordination complexes, and there is much excitement at the prospect that single-atom catalysts can be used to "heterogenize" problematic reactions currently performed in solution (*9-11*). Although there are similarities, there are also practical differences. Homogeneous catalysts are designed for purpose based on a fundamental understanding of the structure-function-relationship, and the ligands play an important role activating reactants and stabilizing intermediates. Complexes containing $O^{2-}$ ligands are rare (*5*), and the binding environment of the metal adatom on the metal oxide is difficult to ascertain and control. Because robust, inexpensive metal oxides are set to continue as the support of choice in single-atom catalysis, it is vital to learn how the coordination of the metal site on an oxide surface affects its adsorption properties and ultimately catalytic activity.

Most studies of oxide-supported single-atom catalysts feature transmission electron microscopy (TEM) images (*12*) of powder samples showing that the metal adatoms align with the cationic sublattice. These structures are in agreement with x-ray absorption near edge structure (XANES) spectra, which typically suggest coordination to oxygen. Density functional theory (DFT) calculations (*13-16*), also predict such sites to be most stable (albeit on simplified models of the support surface). Often, a positive charge state can be inferred from x-ray photoelectron spectroscopy, infrared vibrational spectroscopy, or both of adsorbed CO (*8, 13*). More definitive interpretation requires simulated or experimental reference spectra, and thus a knowledge of the atomic configuration around the active site (*17*). Nevertheless, the charge state of the metal is frequently invoked to explain reactivity, although this explanation is not without controversy. The cationic nature of $Pt_1$ has been linked to both stronger (*8*) and weaker (*15*) CO adsorption, with diametrically opposed conclusions regarding CO oxidation activity.

In this paper, we combine temperature-programmed desorption, x-ray photoelectron spectroscopy, scanning tunneling microscopy, and DFT calculations to study CO adsorption on a series of model single-atom catalysts: $Cu_1$, $Ag_1$, $Au_1$, $Ni_1$, $Pd_1$, $Pt_1$, $Rh_1$, and $Ir_1$ atoms on $Fe_3O_4(001)$. Upon deposition at room temperature, all of the metals assume the same twofold coordination to surface oxygen, which allows for direct comparison. We selected CO as the probe molecule because of the abundance of experimental and computational results on metal surfaces and nanoparticles and because many single-atom catalysis studies focus on CO oxidation and the water gas shift reaction, where CO is a reactant. Moreover, the stretching frequency of adsorbed CO is often used to probe the charge state of the metal in single-atom catalysis systems (*8, 13, 17, 18*), which makes a detailed understanding of the metal-CO interaction particularly important.

Our results show that the CO binding strength broadly follows the trends established for the corresponding low index metal surfaces, i.e. group 9 > 10 > 11, but differences emerge within the groups. Our DFT-based calculations reproduce the experimental data well, and show that the adsorption energy is linked to both the electronic structure of the adatom, and to adsorbate-induced structural relaxations of the system. The combination of the two effects differs from metal to metal because each responds differently to the twofold coordination environment, and because each has a different relative affinity for CO and O. Where the agreement between experiment and theory is imperfect, we show that the catalyst structure evolves during the experiment. Ultimately, our results corroborate that the behavior of single-atom catalysts is better rationalized by analogy to coordination complexes rather than metal nanoparticles, and crucially, that the reactivity of the metal atoms in single-atom catalysis can be tuned if the coordination environment can be controlled.

The experiments described here were performed on several natural $Fe_3O_4(001)$ single crystals over a period of 4 years. The samples were prepared in ultrahigh vacuum by cycles of $Ar^+$ or $Ne^+$ sputtering and 900 K annealing. The resulting surfaces exhibited the $(\sqrt{2}\times\sqrt{2})R45°$ low-electron energy diffraction pattern and scanning tunneling microscopy signature characteristic of the so-called subsurface cation vacancy (SCV) reconstruction (*19, 20*), which is known to stabilize dense arrays of metal atoms to temperatures as high as 700 K (*20, 21*). Metal was evaporated directly onto the as-prepared $Fe_3O_4(001)$ support in ultrahigh vacuum (i.e. no additional ligands are present). We define coverages in monolayers (ML, which corresponds to 1 atom per ($\sqrt{2} \times \sqrt{2}$)R45° unit cell, or $1.42 \times 10^{14}$ per cm²). In Fig. 1A, scanning tunneling microscopy shows that Cu, Ag, Au, Ni, Pd, Pt, Rh, and Ir atoms all adsorb at the same location between the underlying rows of surface Fe atoms. Note that the surface oxygen atoms are not imaged in scanning tunneling microscopy because they have no electronic states in the vicinity of the Fermi level, but their position is well known from quantitative electron (*19*) and x-ray diffraction (*22*) measurements and DFT-based calculations. The as-deposited structure and thermal stability of the adatom systems shown in Fig. 1A have all been thoroughly characterized previously (*21, 23-30*), which makes this an ideal model system to systematically compare the adatom properties. The adsorption site is twofold coordinated to the surface oxygen atoms that do not have a subsurface tetrahedral Fe neighbor (Cu is shown as an example in Fig. 1B). This site is essentially where the next Fe cation would reside if the bulk structure continued outward, and the nearest neighbor distance of these sites is the periodicity of the surface reconstruction (8.4 Å). The height of the adatoms above the surface (*z*) varies greatly, and we recently measured this parameter for Cu, Ag, and Ni (*27, 28*).

These data were used to benchmark our theoretical approach, and Fig. 1A shows results from calculations using the optimal DFT+U parameters ($U_{eff}$ = 3.61 eV, optB88-DF functional) (*27, 28*). All of the metal adatoms are cationic with Bader charges in the range 0.28 to 0.68e. Although the Bader derived charges generally underestimate the formal oxidation states, they are a useful scheme to correlate effective and ionic charges (*17, 31*). For example, the Ni adatom exhibited a Bader charge (+0.68e) midway between the metal (+0.01e) and a fivefold coordinated cation at the NiO(100) surface (+1.19e). Because the cations in NiO are nominally 2+, we assign a 1+ oxidation state for the $Ni_1/Fe_3O_4(001)$ adatom. The charge transferred from the metal atom is donated to the undercoordinated surface oxygen atoms, which have a small magnetic moment on the clean surface and are nominally $O^{1-}$ (*19*). These revert to zero magnetic moment and a $O^{2-}$ state when their fourfold environment is recovered. Ag, Au, and Pd are rather weakly bound ($E_{ad}$ = -2.75, -2.90, and -2.99 eV, respectively), whereas Ir shows the strongest interaction ($E_{ad}$ = -5.26 eV). Cu, Ni, Rh, and Pt lie in between -3.98 and -4.60 eV. Thus, the binding energy scales somewhat with the oxophilicity of the metal, in line with previous reports for oxide-supported metal clusters (*32*) and single-atom catalysts (*33*). We note that Ni (*25*), Rh (*30*), and Ir (*29*) adatoms were ultimately unstable against incorporation into the surface lattice, which is important in interpreting our results.

To determine how strongly CO binds to the different metal adatoms, we conducted a series of temperature-programmed desorption experiments. The interaction of CO with the $Fe_3O_4(001)$ support is weak (*34*) and it desorbs from surface $Fe^{3+}$ sites in two peaks between 60 K and 100 K (Fig. S1). Additional small desorption peaks between 100 K and 220 K arise from CO desorbing from $Fe^{2+}$-containing defects in the surface, such as antiphase domain boundaries and step edges (*34*). The gray curves in Figs. 2, A to G, show the "clean-surface" CO temperature-programmed desorption data acquired prior to adsorption of the metal adatoms (omitting the much larger desorption peaks from the regular Fe sites, Fig. S1). The small differences in the relative intensity of the different defect peaks from experiment to experiment were the result of these data being acquired with several different $Fe_3O_4(001)$ samples.

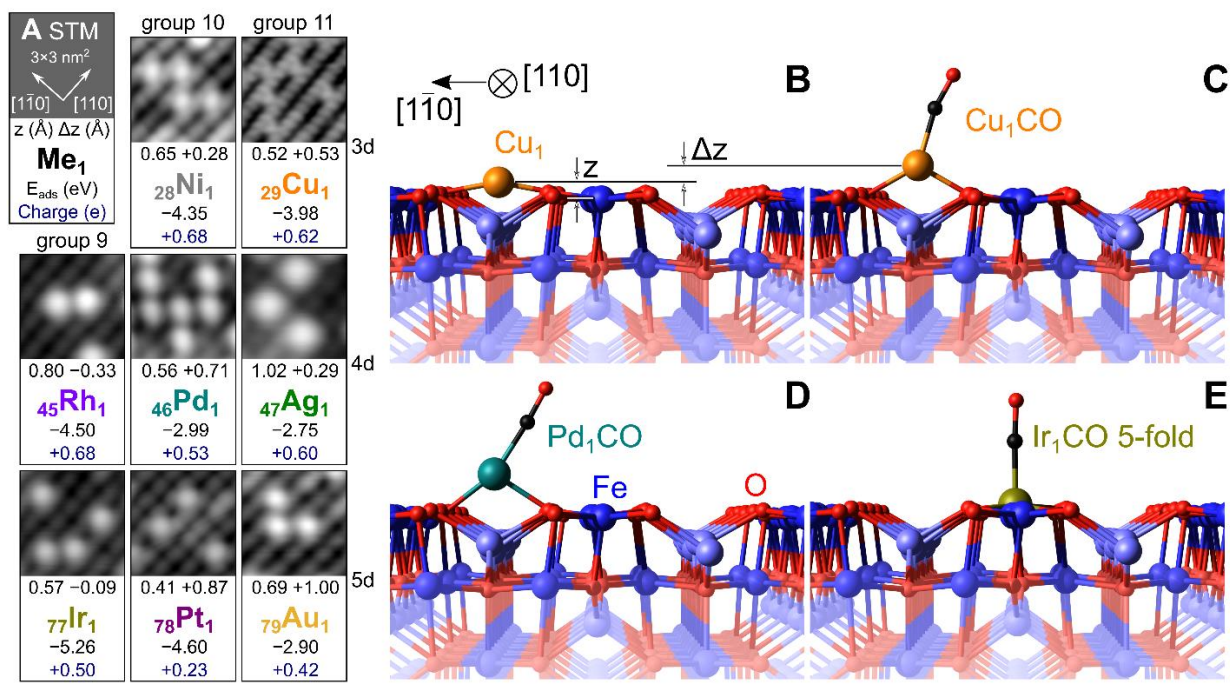

**Fig. 1**: **Characterization of the Fe₃O₄(001)-based single-atom catalysis model systems.** (A) Representative scanning tunneling microscopy images ($V_{sample}$ = + 1-1.5 V, $I_{tunnel}$ = 0.1-0.3 nA) showing metal adatoms adsorbed midway between the bright Fe rows of the Fe₃O₄(001) support. This corresponds to the 2-fold adsorption geometry. Note that the nearest neighbor distance is the periodicity of the surface reconstruction (8.4 Å). Alongside each image are the DFT+U derived adsorption energies, Bader charges, and heights of the $Me_1$ adatom ($z$) above the surface Fe atoms in the 2-fold adsorption geometry, as well as the CO-induced vertical displacement ($\Delta z$). (B-D) DFT+U derived minimum-energy structure for the 2-fold coordinated $Cu_1$/Fe₃O₄(001) adatom before (B) and after (C) adsorption of CO, as well as the $Pd_1CO$ carbonyl (D). Both are lifted from the surface upon CO adsorption. (E) IrCO replaces a 5-fold coordinated surface Fe atom during the temperature-programmed desorption ramp, meaning that CO desorption ultimately occurs from the depicted 5-fold $Ir_1$ geometry (29).

The colored curves in Fig. 2 show selected CO-temperature-programmed desorption data obtained after deposition of metal adatoms at 300 K. In each case, the sample was cooled to 100 K, CO was adsorbed, and the sample heated with a 1 K/s ramp. An arrow marks the temperature-programmed desorption peak corresponding to desorption of CO from the metal adatom in each case. The peak assignments were the result of a series of scanning tunneling microscopy, x-ray photoelectron spectroscopy and temperature-programmed desorption experiments for different adatom coverages and experimental conditions. This assignment was necessary because the $Me_1$/Fe₃O₄(001) systems were not static and evolved differently upon CO exposure and heating. We briefly summarize the basis of the assignment for each metal. Further details can be found in the Supporting Information, and an exhaustive account of these experiments is contained within the PhD thesis of Hulva (35).

Ag and Cu were straightforward to interpret because scanning tunneling microscopy measurements showed the adatom phase was stable for densities up to almost 0.5 ML (27).

Moreover, the adatom arrays were thermally stable to temperatures as high as 700 K (*24*) as well as after exposure to CO. For all adatoms studied here, CO adsorption induced a positive core level shift in the photoemission spectra associated with the adatom (Fig. S2) that disappeared when CO desorbed from the adatom, along with the peak in C1s from the adsorbed CO (Fig. S3).

For both Cu and Ag, a small shoulder was present on the high-temperature side of the main temperature-programmed desorption peak. This shoulder resulted from a fraction of adatoms occupying a metastable geometry after room-temperature deposition and decreased in intensity after annealing the system prior to CO adsorption (See Figs. S4 and S5). For Cu, an additional sharp peak was observed at 120 K that could arise from $Cu(CO)_2$ dicarbonyls, which can be stabilized at low temperature. Details regarding dicarbonyl formation, including how the structures mimic common coordination complexes, can be found in the Supporting Information (Table S1).

Au was the most complicated system studied here because clusters coexisted with adatoms even at very low coverages (<0.15 ML (*21*)), and because two CO temperature-programmed desorption peaks grew together at 335 K and 285 to 300 K as the Au coverage was increased (Fig. S6). These results suggest the presence of two inequivalent sites for adsorption. We assigned the peak at 285-300 K to CO desorbing from regular Au adatoms because pre-annealing the system prior to CO adsorption increased the intensity of the 300 K peak relative to the 335 K peak, and because the 335 K peak intensity increased after exposure to water (Fig. S7). The latter observation suggested the 335 K peak was probably related to Au adatoms interacting with surface hydroxyl groups, as has been observed previously by scanning tunneling microscopy for Pd adatoms on this surface (*23*).

Turning to the group 10 metals, the CO peak desorbing from $Ni_1$ at 300 K was straightforward to assign based on the CO-induced core level shift. The additional peak at 200 K may be related to Ni incorporated in the surface, as Ni can move to subsurface sites at room temperature (*28*). temperature-programmed desorption data for Pd is not shown because CO adsorption destabilized the adatoms and led to rapid agglomeration (*23*). A similar process occurred for Pt, but the temperature-programmed desorption data show a main peak that resulted from the decomposition of $Pt_2(CO)_2$ species (*26*).

The CO desorption peaks from the group 9 Rh and Ir adatoms are again straightforward to assign, although these metals incorporated into the oxide lattice when CO desorbed (*29*). Thus, the post-desorption x-ray photoelectron spectroscopy peaks of Rh and Ir exhibited a higher binding energy than the initial twofold adatom (*29*), rather than shifting back to the as-deposited position. The data acquired at 0.1 ML Rh coverage is shown because this metal exhibited a lower-temperature desorption peak for coverages above 0.2 ML (*35*).

In Figure 2H, we convert the temperature-programmed desorption peak temperatures into desorption energies and plot these alongside the corresponding (111) and (110) metal surfaces obtained from Table S6 and ref (*36*). The basic trend is somewhat similar to the adatoms studied here (i.e. group 9 > 10 > 11), but there are large differences in the absolute magnitudes as well as in the tendencies within groups. It is important to note that most studies on extended metal surfaces typically utilize the Redhead (*37*) or Polanyi–Wigner equation with a pre-exponential factor of about $10^{13}$ s$^{-1}$, which corresponds to a molecule adsorbed in a corrugation-free surface potential. In our case however, scanning tunneling microscopy experiments before and after CO desorption reveals no cluster formation, which suggests the CO molecule is constrained at an immobile adatom even at the desorption temperature. We thus treat the system as an ideal 2D lattice gas (*38*). The low density of adsorption sites results in much higher pre-exponential factors, and assuming the adsorbed CO possesses three hindered rotational modes with a frequency of 400 cm$^{-1}$, we obtain values between $1.27\times10^{19}$ s$^{-1}$ for Cu and $3.25\times10^{19}$ s$^{-1}$ for Ir. The upper and lower bounds for the experimental energies include a temperature uncertainty of $\pm$ 10 K ($\pm$ 20 K for Au).

Figure 2H also includes the adsorption energies determined from DFT+U for a single CO molecule adsorbed on the metal adatoms (filled circles, also see Table S1). The calculations track the experimental trend with a positive offset of $\approx$+0.2 eV for all metals (other than Pt, which formed dimers), which shows that the optB88-DF functional systematically overbind the CO molecule (as suggested previously for water adsorption on Fe$_3$O$_4$) (*39*). Nevertheless, the agreement is excellent, which gave us confidence to delve into the details of the calculations and understand how the "single atom" nature of the systems affects CO binding.

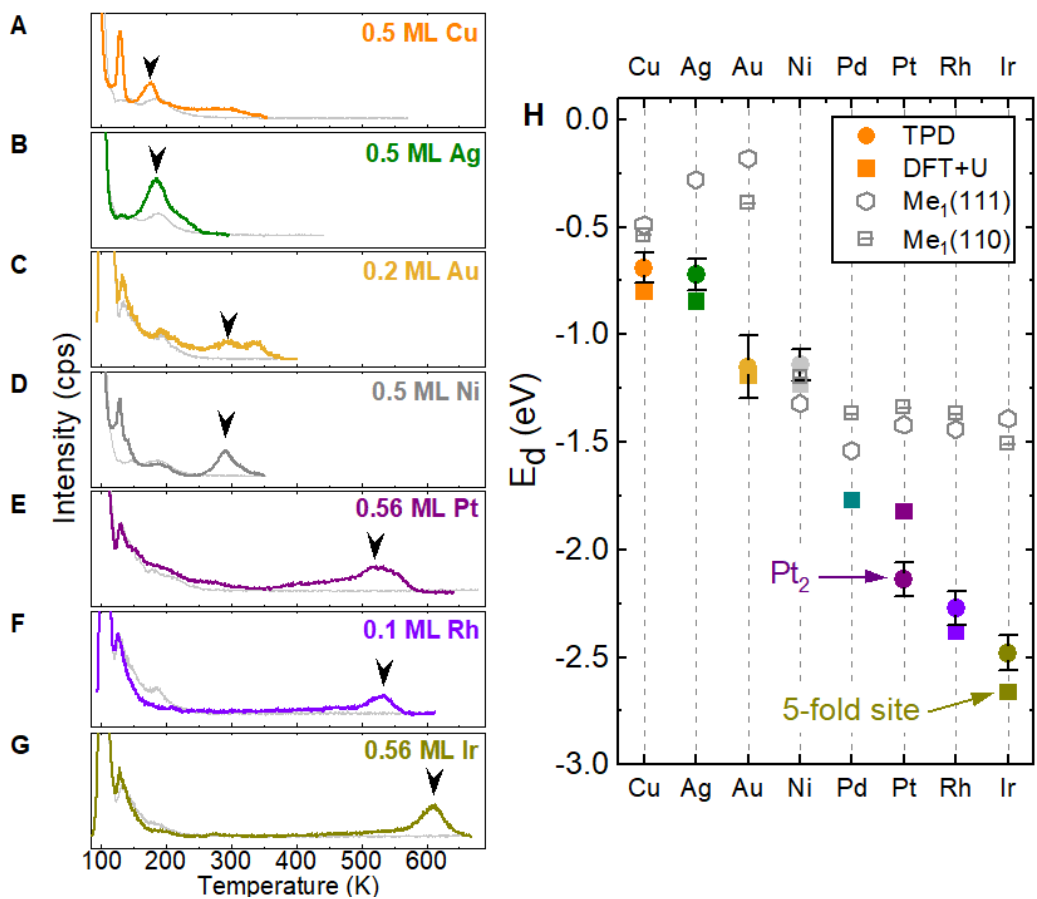

**Fig. 2**: **Determining the CO adsorption on Me$_1$/Fe$_3$O$_4$(001) model catalysts with temperature programmed desorption**. (A-G) CO temperature-programmed desorption curves for various adatoms (1 monolayer (ML) corresponds to 1 metal atom per surface unit cell, or $1.42\times10^{14}$ atoms per cm$^2$). Desorption peaks associated with the metal adatoms are labelled using an arrow. Reasoning behind their position is given in the main text. The light gray curves show CO temperature-programmed desorption data acquired for the ultrahigh vacuum-prepared

$Fe_3O_4(001)$ surface prior to deposition of the metal. (H) Plot of experimental and calculated CO adsorption/desorption energies, alongside experimental values for respective metal (111) and (110) surfaces taken from the literature (see Table S6).

The d-band center, d-band center of mass and the d-band filling are well-known descriptors affecting the CO adsorption energy on transition metal surfaces (*1, 36, 40*). When the outermost d shell is full, as for group 11 metals, CO binding is dominated by d-2π* back-donation. When the d shell is partially empty, as for group 9 metals, 5σ-d donation plays a more important role, and even dominates in the case of Ir (*41*). Placing a metal atom on a metal oxide support leads to the formation of chemical bonds, which affects the symmetry of the d states as well as the filling. On this basis, one would expect the CO adsorption energy to differ from site to site and metal to metal, and that there might be parallels with metal oxide surfaces.

To understand how the nature of the single-atom catalysis site affects the CO adsorption energy, an extensive set of calculations was performed in which CO adsorption on the $Me_1/Fe_3O_4(001)$ systems was compared to CO in atop sites (i) on the corresponding (111) metal surfaces and (ii) on surface cation sites at the most stable facets of the respective metal oxide. The full dataset, all acquired using the same computational setup, is shown and discussed in the Supporting Information. In what follows, we discuss three representative metals, Ni, Ag, and Ir, which allow us to illustrate the most important factors involved. We found that the reactivity of metals was governed by electronic structure, but that CO-induced structural distortions modulate the CO adsorption energy, particularly when the coordination is low.

It would be reasonable to expect that the oxide-supported metal adatoms should have properties somewhat in between a pure metal and a metal oxide. This case was exemplified by the Ni adatom, which exhibited a Bader charge (+0.68e) midway between the metal surface (+0.01e) and a fivefold coordinated cation at the NiO(100) surface (+1.19e). Because the cations in NiO are nominally 2+, we assigned a 1+ oxidation state for the $Ni_1/Fe_3O_4(001)$ adatom. $Ni^{1+}$ is not common in nature, which immediately shows that unusual properties can arise when metal atoms are stabilized in undercoordinated geometries. In Fig. 3A, we plot the CO adsorption energy ($E_{ads}(CO)$) as a function of d-band center of mass for the three different environments mentioned above. As the metal atom becomes more oxidized (Bader charge increased), the d-band center of mass moves further below the Fermi energy $E_F$, and the CO binding weakens. If the Ni atom replaces a fivefold coordinated Fe at the $Fe_3O_4(001)$ surface, as has been observed in experiment (*42*), the system resembles NiO(100) because the local environment is almost identical. In this configuration, the d-band center of mass is far away from $E_F$ and d-2π* back-donation cannot occur effectively, so the CO-Ni bond is primarily electrostatic (*43*).

In the case $Ag_1/Fe_3O_4(001)$ (Fig. 3B), CO is bound much more strongly than for the corresponding metal surface because the d-band center of mass shifted by 1.1 eV toward $E_F$ in the twofold-coordinated adatom site. Because Ag retains a filled d-shell in the 1+ state, d-2π* back-donation continues to dominate, and the CO bond is substantially strengthened. However, the interaction is so strong that the $Me_1$-support bonds are weakened to accommodate the molecule in a more favorable geometry. Our reference calculations for the $Ag_2O(111)$ surface suggest that the CO adsorption energy is optimal when the $O-Me_1$-CO bond angle is 180° (see Supporting Information). Such an ideal geometry cannot be achieved at the $Fe_3O_4(001)$ lattice, and the system comes to equilibrium when the energetic gain achieved by strengthening the $Me_1$-

CO bond balances the cost of weakening the Me₁-O bonds. Thus, the adsorption energy is weaker than what would be expected based on electronic structure of the adatom alone.

A similar distortion of the local structure was observed for all group 10 and 11 metals. In the $Cu_1/Fe_3O_4(001)$ system, the adsorption energy was almost exactly the same as calculated for the metal surface. This situation is a coincidence, however, because the stronger bond caused by the upward shift in the d-band center of mass was almost exactly compensated by the large CO-induced distortion (see Fig. 1, B and C). The most extreme consequence of CO-induced distortion was the sintering observed for Pd and Pt. Here, the formation of a stable carbonyl weakened the support interaction to such an extent that diffusion becomes facile even at room temperature.

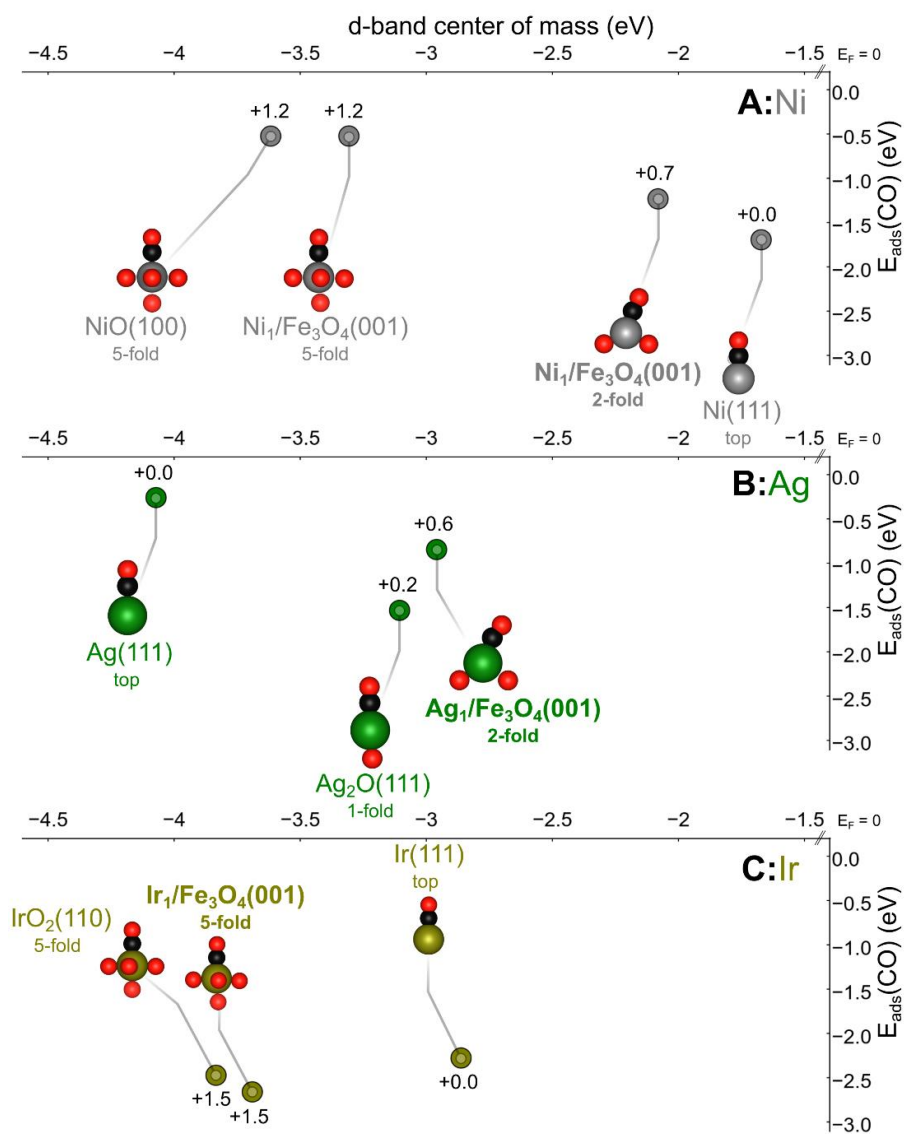

**Fig. 3: CO adsorption energy for Me₁/Fe₃O₄(001) single-atom catalysts compared with Me(111) and MeOₓ oxide surfaces, plotted against the d-band center of mass.** The first coordination sphere of each configuration is illustrated with oxygen red and carbon black, and

each point is labelled with the corresponding Bader charge. (A) Nickel: As the oxidation state increases, the d-band center of mass moves away from $E_F$ and the CO binding energy decreases. Ni incorporated into a fivefold site on $Fe_3O_4(001)$ is close to $NiO(100)$ due to the similar bonding environment. (B) Silver: Oxidized Ag has its d-band center of mass shifted toward $E_F$ leading to a stronger CO binding energy compared to the metal. The onefold coordinated Ag atom on $Ag_2O(111)$ exhibits a strong $E_{ads}(CO)$ due to a linear $O_{surface}$-Ag-CO geometry. (C) Iridium: The $E_{ads}(CO)$ at 5-fold $Ir_1/Fe_3O_4(001)$ is enhanced over $Ir(111)$ due to strong oxidation of the cation, which enhances $5\sigma$-d donation. In this site, the Ir cation is similar to a cation in the $IrO_2(110)$ surface.

Finally, Ir exhibited the strongest CO binding of all the systems considered here. Upon CO adsorption, the twofold coordinated Ir adatom formed a bond to a subsurface O atom and created a pseudo square-planar environment for the Ir atom (29). Binding a second CO molecule created a highly stable $Ir(CO)_2$ dicarbonyl with two bonds to surface oxygen (29), again highlighting the preference for the square-planar coordination. Dicarbonyls can be formed on almost metals studied here, and the structure follows that expected on the basis of Me(I) coordination complexes (see Table S3 and associated discussion).

In any case, the twofold Ir atom is ultimately unstable against incorporation into the $Fe_3O_4(001)$ lattice and replaces a fivefold coordinated Fe cation at ~ 450 K. Thus, CO desorption occurs from a fivefold $Ir_1/Fe_3O_4(001)$ site in the temperature-programmed desorption experiment. Because the Ir-CO bonding is dominated by $5\sigma$-d donation (41), the high oxidation state of the Ir atom (Bader charge = 1.5) in this coordination enhances the CO adsorption energy. The downward shift in d-band center of mass weakens d-$2\pi^*$ back donation, but this is counteracted by the better alignment of the $5\sigma$ and d levels. Little distortion occurs when CO completes the favored sixfold coordination environment (Fig. 2D), so the adsorption energy is purely based on the electronic structure. Indeed, the fivefold $Ir_1/Fe_3O_4(001)$ site closely resembles a fivefold Ir cation at the rutile $IrO_2(110)$ surface in terms of Bader charge, location of the d-band center of mass, and $E_{ads}(CO)$. This result suggests that fivefold Ir single atom catalysts could be cost-effective replacements for $IrO_2$ electrocatalyts for water splitting, which are highly effective but prohibitively expensive.

Whether a metal incorporates into the $Fe_3O_4$ support is governed by the oxygen affinity and preferential O coordination. Table S7 lists the overall energetic gains moving from a two- to a fivefold site. The group 11 elements will not incorporate because they prefer the twofold coordination in their bulk oxide, and would be over-saturated in a fivefold geometry. The group 10 elements Pd and Pt prefer to be fourfold coordinated to oxygen, and have similar energies for two and fivefold geometries studied here. In any case, neither is competitive against sintering into clusters, so incorporation does not occur. Finally, the group 9 elements as well as Ni prefer incorporation into the lattice because they have a high oxygen affinity and have octahedral coordination in their most stable oxide.

The implications of our results for single-atom catalysis are twofold. The choice of metal for a specific reaction is generally governed by the Sabatier principle, which states that the optimum catalyst should be reactive enough to adsorb and activate reactants, but not so reactive that the rate is hampered by sluggish desorption of products. Single-atom catalysis was originally developed because expensive metals such as Pt exhibit the optimal balance, and sit at the top of a "volcano plot". Our study shows that binding the metal atoms to an oxide alters their electronic structure. For most metals, the twofold binding site strongly shifted the d-band center of mass toward $E_F$ and increased the metal-CO bond strength. This resulted in the shorter Me-C bond lengths compared to the metal surface (Table S4). Moreover, because adsorption energies are a good proxy for activation barriers (*44*), we conclude that the metal adatoms become more reactive, so the metals previously limited by poor reactivity will shift closer to the top of a volcano plot. Note that we refer to bond strength here, not adsorption energy, because the structural distortions render these two quantities inequivalent for the twofold coordinated structures.

This result also has consequences for the catalysts limited by product desorption because destabilization by structural distortions will lead to product desorption at lower temperatures, that is, avoid poisoning by adsorbed products. Stabilizing low-coordination "single-atom" sites could provide new ways to tailor the activity of specific metals and allow non Pt-group metals to become competitive. This conclusion is in line with two recent theoretical screening studies (*14, 16*), which suggest that $FeO_x$-supported $Pd_1$, $Rh_1$, $Ru_1$, and even $Ti_1$ or $Co_1$ species can have similar activity as $Pt_1$ for CO oxidation.

Knowledge of the local geometry is thus essential to understand the adsorption properties of supported metal adatoms. The observed properties stem directly from how the particular metal reacts to a particular bonding environment, and we observed a preference for the systems to adopt the coordination and bond angles seen in corresponding coordination complexes when CO is adsorbed. This, coupled to the observed ability of the single-atom catalysis systems to distort the structure to accommodate reactants, suggests that adsorbed single atoms should be viewed as undercoordinated complexes, and not as the smallest possible nanoparticle. The results seen here will be transferrable to many oxide supports, including zeolites, which also offer a twofold coordination environment. Particularly stable configurations are possible if the metal/oxide system forms solid solution in the bulk (e.g. the fivefold $Ni/Fe_3O_4(001)$), in which case the metal site is straightforward to predict. Although we observe a clear driving force for higher coordination (incorporation) for many of the metals studied here, low coordination to the oxide could be stabilized by the adsorption of additional ligands (reactants, OH, water) in the reactive environment, provided these create a stable geometry for the metal atom. If control of the active site can be achieved in industrial systems, single-atom catalysts really do have the potential to combine the best of heterogeneous and homogeneous catalysis.

**Acknowledgments:** The authors would like to thank Peter Blaha (TU Wien) and Paul T. P. Ryan (Imperial College London) for useful discussions. **Funding:** GSP, JH, MM, ZJ and RB acknowledge funding from the Austrian Science Foundation (FWF) Start Prize Y847-N20, and UD and FK acknowledge the Austrian Science Fund FWF (Project 'Wittgenstein Prize, Z250-N27). GSP and MM acknowledge funding from the European Research Council (ERC) under the European Union's HORIZON2020 Research and Innovation program (ERC Grant Agreement No. [864628]). ZJ also acknowledges support from the TU Wien Doctoral Colleges TU-D. The computational results were achieved in part using the Vienna Scientific Cluster (VSC 3 and VSC 4) **Author contributions:** JH and RB performed the experiments under the supervision of GSP, who conceptualized the research and acquired research funding for the project. MM performed the theoretical calculations under the supervision of CF. JH, MM and GSP wrote the paper with substantial input/revision from UD, MS, FK, ZJ and CF. **Competing interests:** Authors declare no competing interests; and **Data and materials availability:** All data is available in the main text or the supplementary materials.

**Supplementary Materials:**

Materials and Methods

Computational Details

Figures S1-S14

Tables S1-S7
References (44-95)

# Science

## AAAS

Supplementary Materials for

Unravelling CO Adsorption on Model Single-Atom Catalysts

Jan Hulva, Matthias Meier, Roland Bliem, Zdenek Jakub, Florian Kraushofer, Michael Schmid, Ulrike Diebold, Cesare Franchini, Gareth S. Parkinson

Correspondence to: parkinson@iap.tuwien.ac.at

**This PDF file includes:**

Experimental details
**Figure S1:** Temperature programmed desorption spectra for CO on the as-prepared $Fe_3O_4(001)$ surface.
**Figure S2:** Summary of the XPS core levels of $Cu_1$, $Ag_1$, $Au_1$, $Ni_1$, $Pt_1$, $Rh_1$, and $Ir_1$ adatoms following deposition at room temperature.
**Figure S3:** Summary of the C1s XPS peaks assigned to CO adsorbed on $Cu_1$, $Ag_1$, $Au_1$, $Ni_1$, $Pt_2$, $Rh_1$, and $Ir_1$.
**Figure S4:** Repeated CO TPD for 0.8 ML of $Ag_1/Fe_3O_4(001)$ for consecutive TPD experiments with the ramp terminated at 290 K.
**Figure S5:** Consecutive CO TPD measurements on the $Cu_1/Fe_3O_4(001)$ surface (0.5 ML Cu).
**Figure S6**: Additional CO TPD data for the $Au_1/Fe_3O_4(001)$ system.
**Figure S7:** TPD experiments following co-dosing of CO and $D_2O$ on $Au_1/Fe_3O_4(001)$.
**Computational details**
**Table S1:** $U_{eff}$ (= U−J given in eV) applied for the different systems studied here.
**Table S2:** Computational setup.
**Table S3:** DFT results including the adsorption energies of the metal adatom, and the first and second CO molecule.
**Table S4:** Structural parameters for the metal-CO systems determined by PBE+U calculations with and without adsorbed CO.
**Figure S8:** Summary of all calculated CO adsorption energies.
**Figure S9:** Dominant CO binding mechanism for metal (111) sites by element.
**Computational Results for Group 11 (Cu, Ag, Au)**
**Figure S10:** $E_{ads}(CO)$ as a function of d-band center of mass for Cu, Ag and Au.

**Figure S11:** Relaxations occurring at the 2-fold $Ag_1/Fe_3O_4(001)$ site in comparison with the 1- and 2-fold sites on $Ag_2O(111)$.

**Figure S12:** Adsorption energies for O and CO on a Me atom in the gas phase, Me = [Cu, Ag, Au, Ni, Pd, Pt, Co, Rh, Ir].

**Computational Results for Group 10 (Ni, Pd, Pt)**

Figure S13: $E_{ads}(CO)$ as a function of d-band center of mass for Ni, Pd and Pt.

**Computational Results for Group 9 (Rh, Ir)**

Figure S14: $E_{ads}(CO)$ as a function of d-band center of mass for Rh and Ir.

**Table S5**

**:** CO induced core level shift compared between theory any experiment

**Table S6**

**:** Summary of experimental CO binding energies on different metal surfaces.

**Table S7**: Calculation of the energy gained in moving a twofold coordinated adatom to a fivefold coordinated "incorporated" site substituting surface Fe within the $Fe_3O_4(001)$ lattice.

**Experimental details**

The experiments were performed in two ultrahigh vacuum (UHV) systems with base pressures below $10^{-10}$ mbar, using natural $Fe_3O_4$(001) samples purchased from SurfaceNet GmbH. The samples were prepared *in situ* by sputtering (1 keV $Ar^+$ or $Ne^+$, 10 min) and annealing in UHV (930 K, 10 min). Every other annealing step was performed in a partial pressure of $O_2$ ($p_{O2} = 5 \times 10^{-7}$ mbar, 20 min) to prevent reduction of the crystal selvedge. The metals were deposited using Focus e-beam evaporators, and the deposition rate was calibrated by temperature-stabilized quartz crystal microbalances (QCM). One monolayer (ML) is defined as one atom per ($\sqrt{2} \times \sqrt{2}$)R45° unit cell of the $Fe_3O_4$(001), which corresponds to $1.42 \times 10^{14}$ atoms/cm$^2$. The STM data were acquired in a two-vessel UHV chamber consisting of a preparation chamber ($p < 10^{-10}$ mbar) and an analysis chamber ($p < 7 \times 10^{-11}$ mbar) equipped with an Omicron μ-STM operated at room temperature in constant-current mode with electrochemically etched tungsten tips. The chamber further includes commercial XPS, Auger electron spectroscopy and low-energy electron diffraction instruments. The STM images were corrected for distortion and creep of the piezo scanner as described in reference (*45*).

The TPD and XPS data were acquired in a second chamber, custom built to study the surface chemistry of bulk metal oxide samples. The setup is equipped with a liquid-He cryostat, a home-built molecular beam source, a quadrupole mass spectrometer (Hiden HAL 3F PIC), a monchromatized Al/Ag twin-anode X-ray source (Specs XR50M, FOCUS 500), a hemispherical analyzer (Specs Phoibos 150), a low-energy electron diffraction setup (Specs ErLEED), an ion source (Specs IQE 12/38) and a UV source (Specs UVS 10/35). Full details are given in reference (*46*). The samples were mounted on a Ta sample plate using Ta clips, with a thin Au foil placed between the sample plate and the sample to ensure good thermal contact. The sample plate was attached to the liquid-He cryostat, and the temperature was measured by a K-type thermocouple spot-welded on the sample plate. A molecular beam is formed by expansion of 0.27 mbar CO or $^{13}$CO through two differentially pumped stages. This results in a well-defined beam spot on the sample surface with a diameter of 3.35 ± 0.17 mm and a top-hat intensity profile. The TPD spectra were acquired with a 1 K/s heating ramp.

**Figure S1**

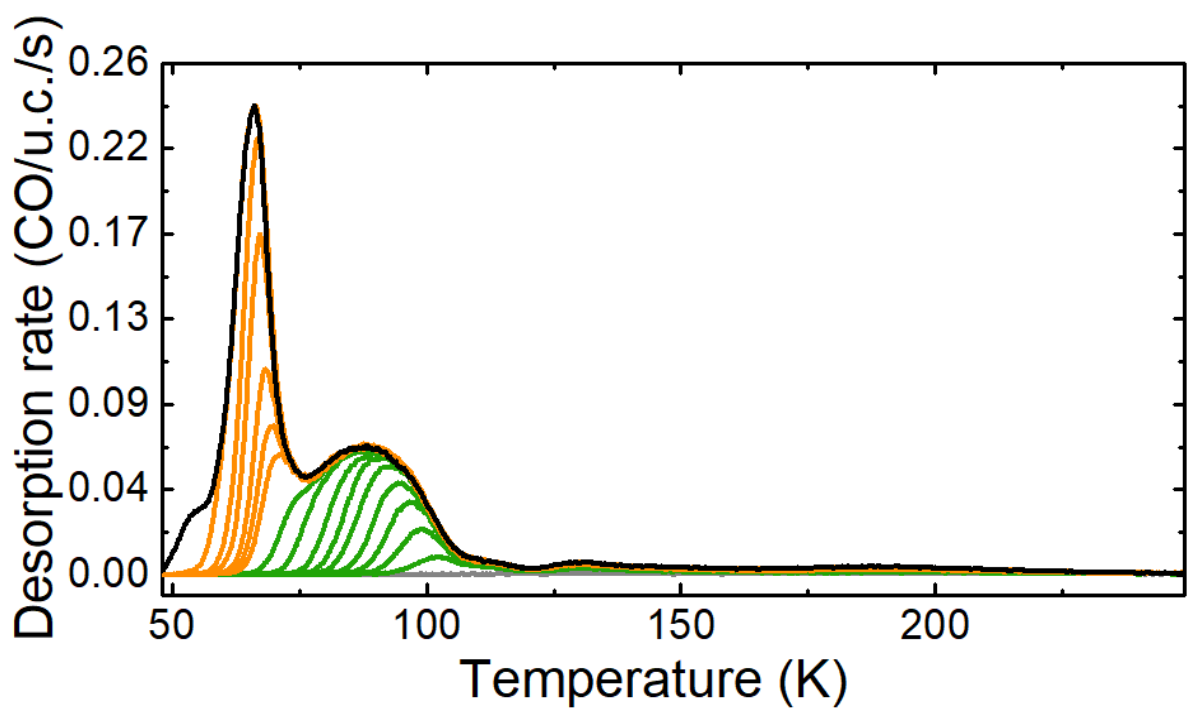

**Figure S1: Temperature-programmed desorption spectra for CO on the as-prepared Fe₃O₄(001) surface.** The green curves are for coverages between 0-2 CO per unit cell, and the orange curves for 2-4 CO per unit cell. The black curve corresponds to saturation of the first monolayer. Note the presence of small defect states at 160 K and 200 K. A full analysis of these data was published in Ref. (*34*).

**Figure S2**

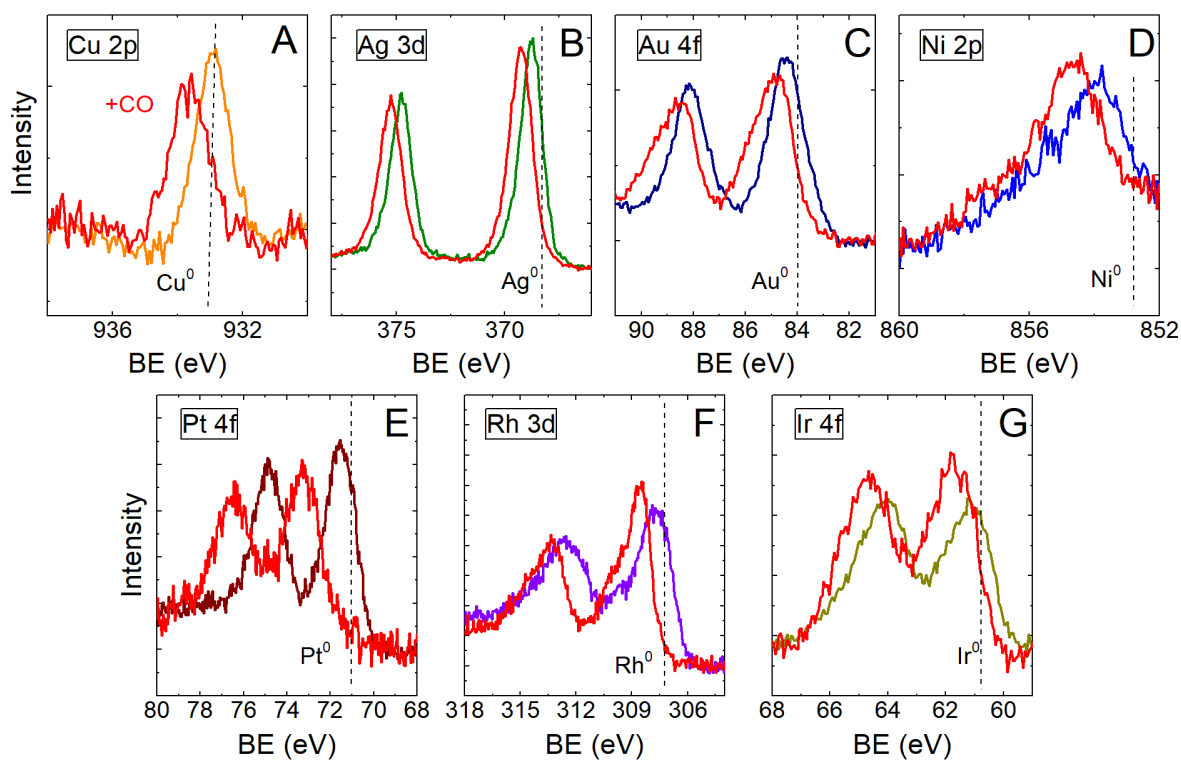

**Figure S2**: **Summary of the XPS core levels of (A) Cu₁, (B) Ag₁, (C) Au₁, (D) Ni₁, (E) Pt₁, (F) Rh₁, and (G) Ir₁ adatoms following deposition at room temperature.** For reference, the dashed, vertical lines mark the positions observed from the corresponding bulk metal. The red curves show the CO-induced shift of the metal core levels after saturating the surface with CO at 120 K. The data were measured by monochromatic Al Kα at an electron emission angle of 80° to the surface normal.

**Figure S3**

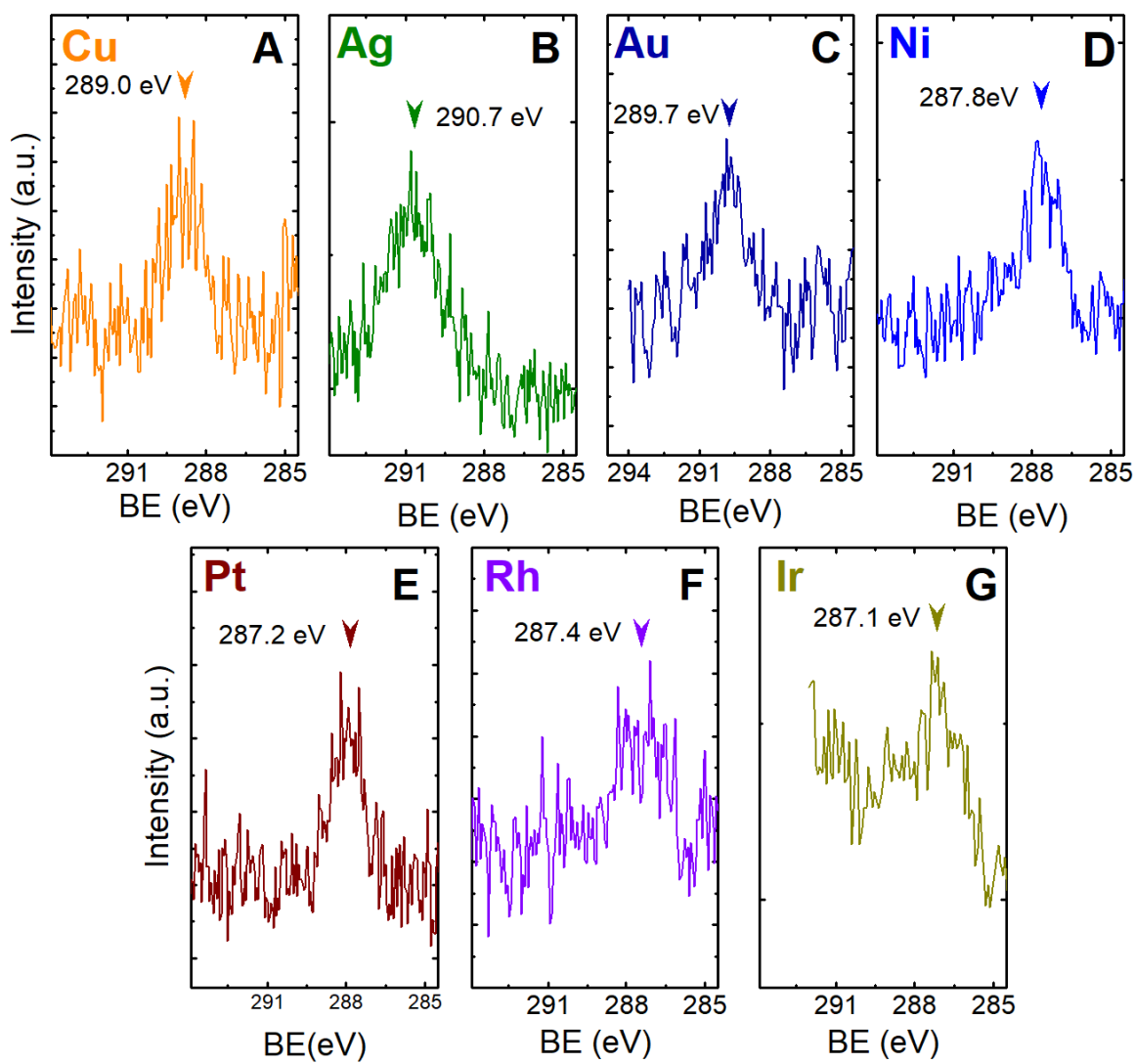

**Figure S3: Summary of the C1s XPS peaks assigned to CO adsorbed on (A) Cu₁, (B) Ag₁, (C) Au₁, (D) Ni₁, (E) Pt₂, (F) Rh₁, and (G) Ir₁.** These data were acquired at the same time as the red curves shown in Figure S2. The data were measured by monochromatic Al Kα at an electron emission angle of 80° to the surface normal. The C1s binding energy decreases systematically as the desorption temperature in TPD increases.

**Figure S4**

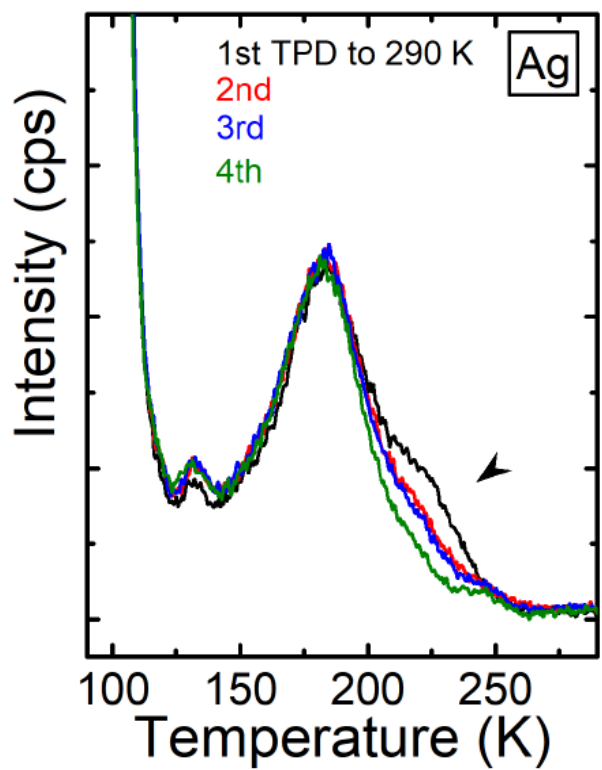

**Figure S4: Repeated CO TPD for 0.8 ML of $Ag_1$/$Fe_3O_4$(001) for consecutive TPD experiments with the ramp terminated at 290 K.** The reduction in the intensity of the shoulder at 220 K is marked by an arrow. These results support the statement (in the main text) that the shoulder is due to adatoms in a metastable geometry.

**Figure S5**

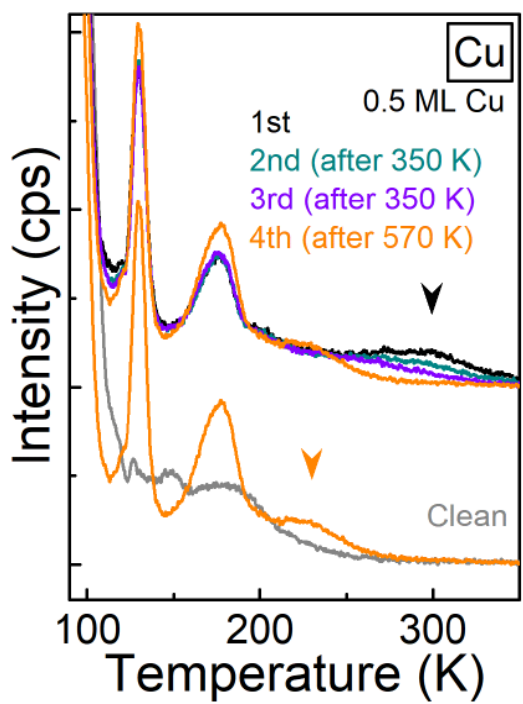

**Figure S5**: **Consecutive CO TPD measurements on the Cu₁/Fe₃O₄(001) surface (0.5 ML Cu).** The intensity of the desorption feature at 300 K (marked by black arrow) is strongest in the 1st TPD experiment after the deposition (black curve) and decreases in the 2nd round (green) and 3rd round (violet). The shoulder disappears in the 4th TPD run (orange), prior to which the sample was heated to 570 K. Both low-temperature desorption features (180 and 130 K) increase in intensity. The intensity of the feature at 230 K slightly increases (marked by the orange arrow). In these experiments, the surface was saturated by CO at 60 K by dosing ~5 CO per Fe₃O₄(001) surface unit cell. The high-temperature shoulder at 300 K is attributed to atoms in a metastable geometry.

**Figure S6**

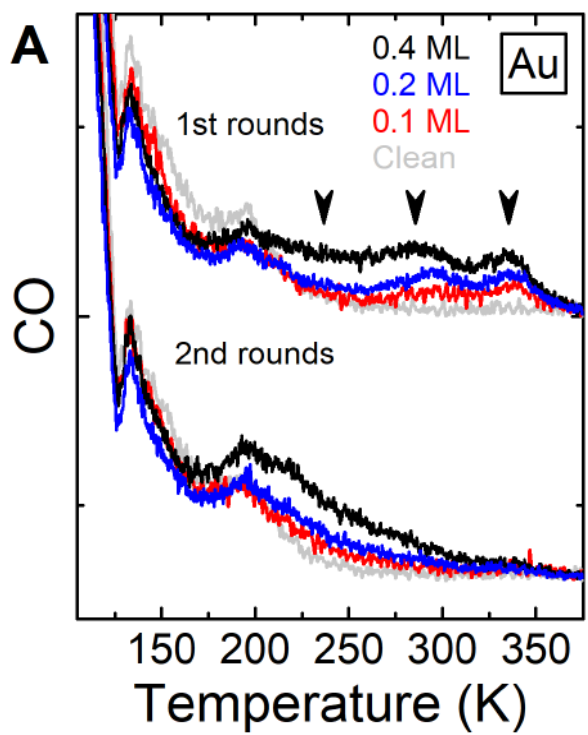

**Figure S6: Additional CO TPD data for the Au₁/Fe₃O₄(001) system.** A) CO TPD for 0.4 ML Au on Fe₃O₄(001) (black), 0.2 ML Au (blue), 0.1 ML Au (red), and the clean surface (grey). Black arrows mark desorption features at 240 K, 285 K, and 335 K. Note that the broad intensity around 240 K grows at the higher coverage, where clusters have been seen to have form in experiment. CO TPD curves in the upper part of the plot are from the first TPD experiment performed for each coverage with an end temperature of the TPD ramp of 400 K. The intensity around 240 K remains in the second round of the TPD, consistent with the idea that clusters are mobile at 400 K, and grow at the expense of the adatom phase. We conclude that the desorption peaks at 285 K and 335 K are linked to adatoms in different environments. The origin of the desorption peak at 345 K is discussed in Fig. S7.

**Figure S7**

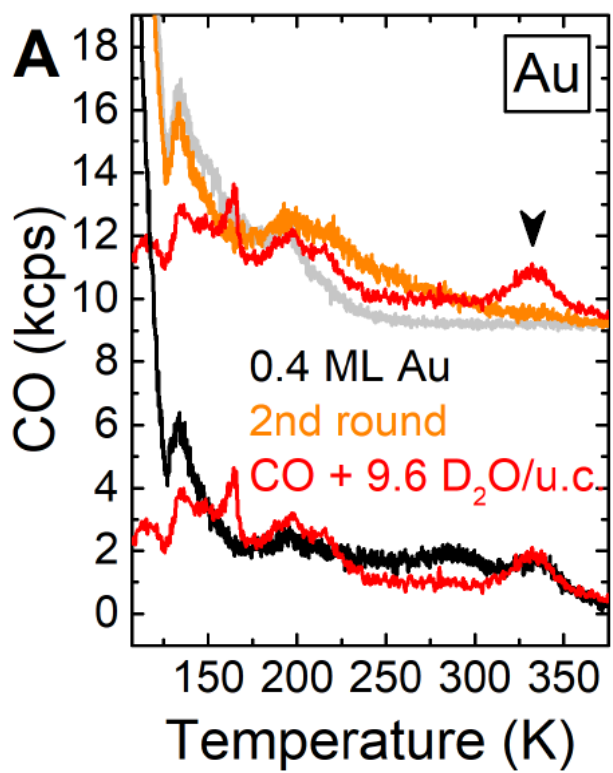

**Figure S7: TPD experiments following co-dosing of CO and D₂O on Au₁/Fe₃O₄(001).**
The red curve shows CO TPD from a 0.4 ML Au/Fe₃O₄(001) sample that was annealed to 400 K, and then sequentially exposed to CO (~3 CO/u.c., saturation) and 9.6 D₂O/u.c. at 90 K (red). The strong peak at 335 K following exposure to water suggests that this peak cannot solely be due to CO desorption from bare Au adatoms. The black curve shows the same CO TPD shown for 0.4 ML Au in Figure S6 (first round), and the orange data shows the "2ⁿᵈ round" TPD obtained after heating the sample to 400 K.

**Computational details**

The Vienna *ab initio* Simulation Package (VASP) (*47, 48*) was utilized for all DFT calculations in this work. The Projector Augmented Wave (PAW) (*49, 50*) method describes the electron and ion interactions, with the plane-wave basis set cut-off energy set to 550 eV. Convergence is achieved when an electronic energy step of $10^{-6}$ eV is obtained and the forces acting on ions become smaller than 0.02 eV/Å. The calculations are spin polarized, and include dipole corrections as implemented in Refs. (*51*) and (*52*).

The Perdew-Burke-Ernzerhof (PBE) (*53*) functional is used for preliminary convergence, with dispersion effects treated by Grimme's D2 (*54*). The calculations were then further optimized using optB88-DF (*55-57*). The pre-convergence is necessary because optB88-DF tends to get trapped in local minima, but in our experience optB88-DF offers more accurate relative energies for small molecules adsorbed on $Fe_3O_4$ surfaces, and is able to treat the magnetite slab properly in terms of its electronic and magnetic properties (*58*). Further details regarding the lattice parameter, the structural model, and the accurately optimized asymmetric slab utilized for the $Fe_3O_4(001)$ surface are discussed extensively in Ref. (*59*).

To understand the data acquired for the $Me_1/Fe_3O_4(001)$ systems, it was necessary to compare the properties to the corresponding pure metal (111) surfaces, as well as surface sites on the most stable metal oxide of the given element. For consistency, and to allow direct comparison of the different systems, the setup described above is used for all such calculations. Homogeneous Γ-centered k-meshes are used for structural relaxations, which are further increased (by ×3) for the calculations of the respective electronic DOS. On-site Coulomb terms (Hubbard terms) $U_{eff}$ (eV) (*60*) are applied as specified in Table S1. Additional details of the setups utilized for the different systems are displayed in Table S2.

All asymmetric slabs are separated by at least 10 Å vacuum in the direction normal to the studied surface. Since it is well known that self-consistent spin moments can be sensitive to the choice of the initial magnetic moments, several calculations have been performed with different initial guesses in order to explore a larger portion of the magnetic configuration space and avoid local minima.

Adsorption energies are given with respect to reference energies calculated with the same $U_{eff}$ and according to eq. (1) and eq. (2) for CO and bare adatoms, respectively.

$$E_{ads}(CO) = E_{surf+Me+CO} - E_{surf+Me} - E_{CO} \qquad (1)$$

$$E_{ads}(Me) = E_{surf+Me} - E_{surf} - E_{Me} \qquad (2)$$

$E_{surf+Me+CO}$ is the energy of the surface with the CO and Me both adsorbed, $E_{surf+Me}$ is the energy of the surface with the Me adatom only, and $E_{surf}$ is the energy of the clean surface. $E_{CO}$ and $E_{Me}$ are the gas-phase reference energies of the CO molecule and Me atom, respectively. The d-band center of mass (d-COM) was calculated using the approach detailed in Refs. (*36*) and (*40*),

$$COM = \frac{\int_{-\infty}^{E_F} \rho(E) \times E \, dE}{\int_{-\infty}^{E_F} \rho(E) \, dE} \qquad (3)$$

where $\rho(E)$ is the electronic density of states, $E$ the energy, and $E_F$ the Fermi energy, which is set to 0. Note that this differs from the d-band center, the descriptor proposed by Hammer and Nørskov (*1*), where empty states are also integrated.

The charge of a given atom is calculated using the Bader approach (*61-63*), where

$$\text{Bader charge} = N_{val.e^-} - N_{Bader\,Vol.\,e^-} \qquad (4)$$

where $N_{val.e^-}$ is the number of valence electrons for a given metal, and $N_{Bader\,Vol.\,e^-}$ is the number of electrons within the Bader volume.

**Table S1**

| Element | $U_{eff}$ (eV) |
|---|---|
| Fe (in Fe$_3$O$_4$(001)) | 3.61 (*58, 64*) |
| Ni (SAC on Fe$_3$O$_4$(001)) | 3.61 |
| Cu (SAC on Fe$_3$O$_4$(001)) | 3.61 |
| Ni (in NiO) | 5.30 (*65*) |
| Cu (in Cu$_2$O) | 4.00 (*66*) |
| Ni (metal) | 0.00 |
| Cu (metal) | 0.00 |

**Table S1:** $U_{eff}$ (= U−J given in eV) applied to the d-electrons in the different systems studied here. For all other systems, no on-site correction was implemented.

**Table S2**

| Systems | Bulk symmetry | Surface periodicity | k-mesh | Slab |
|---------|--------------|---------------------|--------|------|
| $Fe_3O_4(001)$ | $Fd\bar{3}m$ | $(\sqrt{2}\times\sqrt{2})R45°$ | $1\times1\times1$ | 5F + 2R |
| Me(111) | $Fm\bar{3}m$ | $(\sqrt{3}\times\sqrt{3})R30°$ | $5\times5\times1$ | 3F + 2R |
| Me(001) | $Fm\bar{3}m$ | $2\times2$ | $5\times5\times1$ | 3F + 2R |
| $Cu_2O(111)$ | $Pm\bar{3}m$ | $2\times2$ | $3\times3\times1$ | 3F + 3R |
| $Ag_2O(111)$ | $Pm\bar{3}m$ | $2\times2$ | $3\times3\times1$ | 3F + 3R |
| $Au_2O(111)$ | $Pm\bar{3}m$ | $3\times3\times1$ | $2\times2$ | 3F + 3R |
| NiO(001) | $Fm\bar{3}m$ | $2\times2$ | $5\times5\times1$ | 3F + 2R |
| PdO(100) | $P4_2/mmc$ | $2\times1$ | $1\times3\times4$ | 3F + 1R |
| PtO(100) | $P4_2/mmc$ | $2\times1$ | $1\times3\times4$ | 3F + 1R |
| $Rh_2O_3$(r-cut) | $R\bar{3}c$ | $1\times1$ | $4\times4\times1$ | 2F + 2R |
| $IrO_2(110)$ | $P4_2/mnm$ | $1\times2$ | $3\times1\times3$ | 2F + 2R |

**Table S2:** Computational setup. Two-dimensional unit cells (larger than primitive $1\times1$ in order to weaken the CO-CO interaction in neighboring cells), k-meshes and slab sizes (asymmetric setup, F=fixed layers; R=relaxed layers) for the different systems studied. Structures are based on available data taken from the Crystallography Open Database.

**Full Computational Results**

In the main text of this paper, three metals (Ni, Ag, and Ir) were selected to illustrate the behavior contained within a much larger dataset. For completeness, the entire computational dataset of CO adsorption energies is described here. A plot of the CO adsorption energies obtained for all the sites considered in this work is shown in Fig. S8. For each metal, there is one data point (triangle) for CO adsorbed at the 2-fold coordinated adatom at the $Fe_3O_4(001)$ surface, and two data points (square) for metal atoms substituting 5-fold coordinated Fe in the $Fe_3O_4(001)$ surface (if such a site was found to be stable). These two data points differ in whether the displaced Fe is included in a subsurface interstitial site or not, but, ultimately, no substantial differences between these two configurations were observed. In addition, two data points (circles) represent CO adsorbed at atop sites on the (111) and (100) surfaces of the pure metal surfaces. Of these two points, adsorption at (100) surfaces is stronger. Calculations were also performed for a representative metal oxide surface for each element (crosses). In each case, a low-index facet of the most stable metal oxide compound was selected. At the $Cu_2O(111)$ and $Ag_2O(111)$ surface, more than one inequivalent cation site is possible with different coordination/geometry. Stronger CO adsorption occurs at the highly undercoordinated Ag or Cu surface atom; this is also the configuration chosen for $CO/Ag_2O$ in Fig. 3 of the main text.

Essentially, Fig. S8 shows that there is a large spread of CO adsorption energies possible for each element considered, and that the relative position of SAC-type systems, metal, and metal oxide sites also varies. DFT-determined adsorption energies for the different systems and structural parameters obtained from the calculations are tabulated in Tables S3 and S4, respectively.

In what follows, we discuss how CO binds on the different metals by group, starting with the group 11 metals (Cu, Ag and Au). In the course of the discussion, we frequently refer to the results of Ref. (*41*), in which cluster calculations based on an atomic orbital basis set were used to decouple the relative contributions of d-2π* back donation and 5σ-d donation to the CO binding energy for metal (111) surfaces. The results of that work are summarized schematically in Fig. S9. Essentially, back-donation dominates for the noble metals, but donation from the molecule into the d-states becomes more important as one moves down and to the left from Cu in the periodic table. On this basis, one might expect that the increased oxidation state (i.e. decreased d-filling) of a metal atom in a SAC or metal oxide site should increase the relative importance of the 5σ-d donation mechanism to the CO adsorption energy.

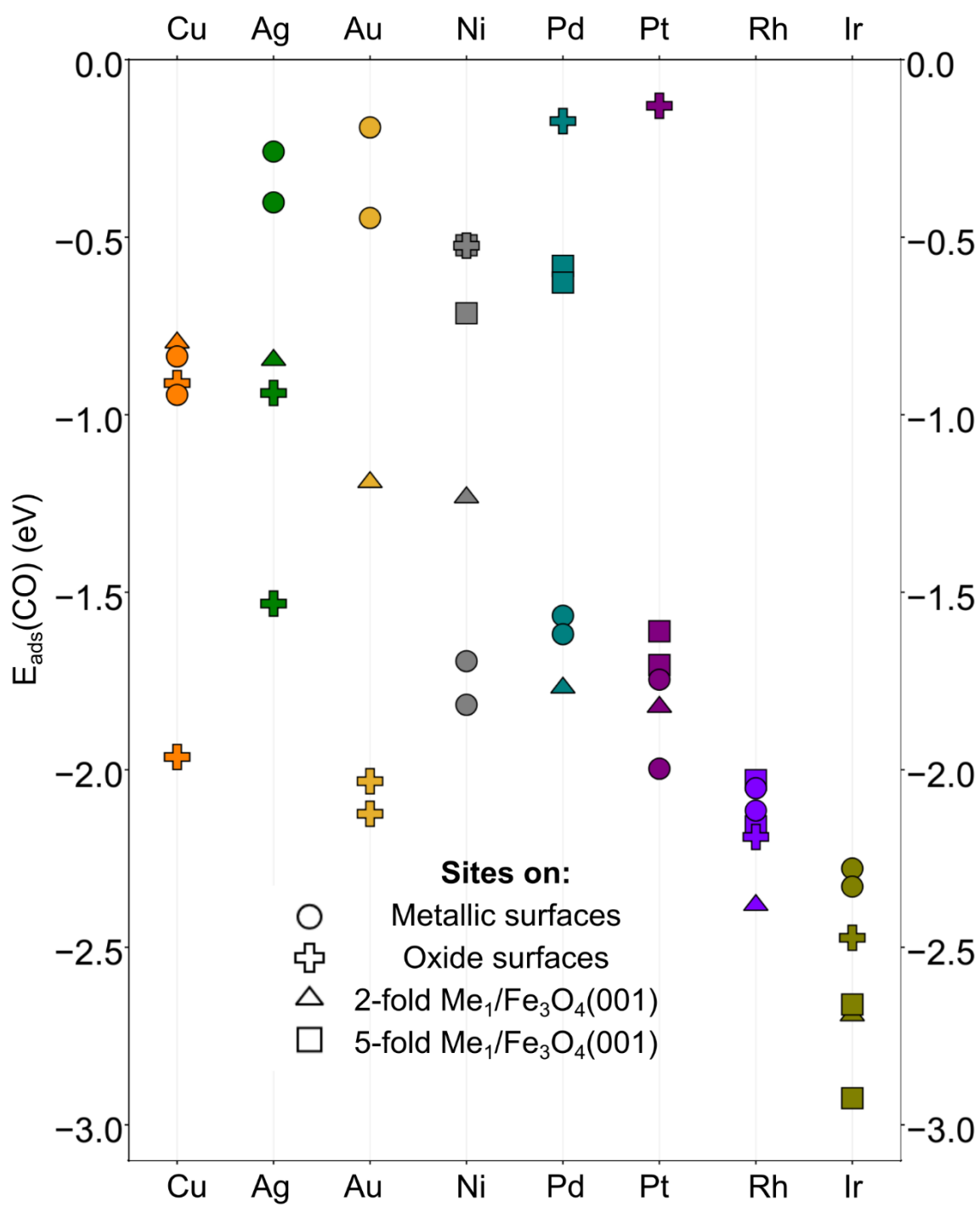

**Figure S8: Summary of all calculated CO adsorption energies.** This includes the (001) and (111) facets of metal surfaces (circles), metal-oxide surfaces (crosses) and SAC sites on $Fe_3O_4(001)$ with different coordination: 2- (triangles) and 5-fold sites (squares).

**Figure S9**

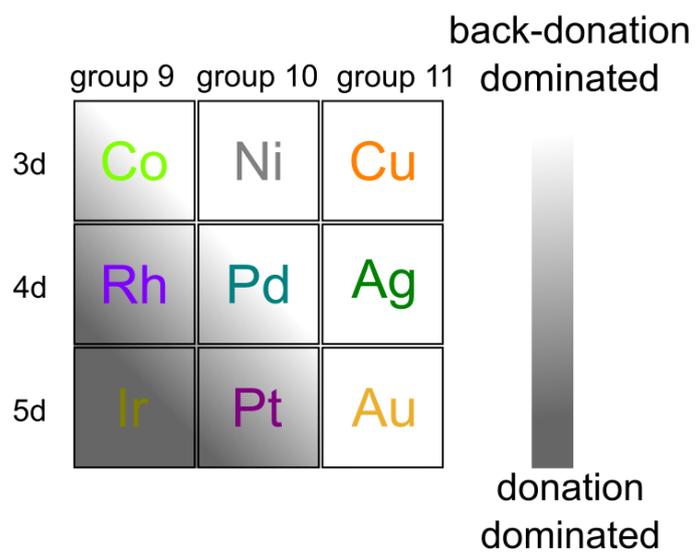

**Figure S9: Dominant CO binding mechanism for metal (111) sites by element, inferred from the data presented in Ref. (*41*).**

**Table S3**

| Metal | Adatom | | Monocarbonyl | | Dicarbonyl | | | (111) surface |
|---|---|---|---|---|---|---|---|---|
| | $E_{ads}(Me_1)$ (eV) | Bader Charge (e) | $E_{ads}(CO)$ (eV) | Bader Charge (e) | $E_{ads}(CO)$ (eV) | $E_{ads}(CO)$ average (eV) | Bader Charge (e) | $E_{ads}(CO)$ (eV) |
| **Cu** | -3.99 | +0.62 | -0.80 | +0.77 | -0.57 (tet) | -0.68 | +0.79 | -0.84 |
| **Ag** | -2.75 | +0.60 | -0.85 | +0.65 | -0.36 (tet) | -0.60 | +0.66 | -0.26 |
| **Au** | -2.90 | +0.42 | -1.19 | +0.52 | -0.07 (tet) | -0.63 | +0.56 | -0.19 |
| **Ni** | -4.35 | +0.68 | -1.23 | +0.74 | -0.55 (tet) | -0.89 | +0.79 | -1.70 |
| **Pd** | -2.99 | +0.53 | -1.77 | +0.58 | -0.55 (sq.pl.) | -1.16 | +0.78 | -1.57 |
| **Pt** | -4.60 | +0.23 | -1.87 | +0.53 | -1.23 (sq.pl.) | -1.53 | +0.88 | -1.75 |
| **Rh** | -4.42 | +0.68 | -2.38 | +0.65 | -1.83 (sq.pl.) | -2.11 | +0.70 | -2.02 |
| **Ir** | -5.26 | +0.50 | -2.69 | +0.83 | -2.21 (sq.pl.) | -2.45 | +0.77 | -2.28 |
| **Ir₁(5-fold)** | -6.23 | +1.49 | -2.79 | +1.45 | --- | --- | --- | --- |

**Table S3: DFT results including the adsorption energies (eV) of the metal adatom, and the first and second CO molecule. The Bader charge of the metal adatom is listed in each case.** Table S3 summarizes the calculated adsorption energies for the metal adatom, and for CO (carbonyl and dicarbonyls). $E_{ads}(CO)$ is given for the second CO molecule with respect to the monocarbonyl, and as an average for both CO molecules. The corresponding Bader charges of the metal atom are given in each case. The dicarbonyls can either produce a tetrahedral or square-planar environment for the metal atom (as previously reported for Ir (*29*)), and the preferred coordination is indicated in the table. For the metals under consideration this agrees with the structure of corresponding Me(I) coordination complex. The TPD peaks observed at 140 K for Ni and Cu appear to be consistent with the desorption of the first CO molecule from the dicarbonyl, which can be formed at the low exposure temperature in our UHV experiments.

**Table S4**

| Metal | | Monocarbonyl | | | Dicarbonyl | | | Metal (111) Surface | | |
|---|---|---|---|---|---|---|---|---|---|---|
| | $z$ | $\Delta z$ | Me$_1$−C | C−O | $\Delta z$ | Me$_1$−C (Average) | C−O (Average) | $\Delta z$ | Me$_1$−C | C−O |
| **Cu** | 0.52 | 0.53 | 1.81 | 1.16 | 0.72 | 1.89 | 1.15 | 0.10 | 1.86 | 1.17 |
| **Ag** | 1.02 | 0.29 | 2.00 | 1.16 | 0.24 | 2.06 | 1.16 | 0.01 | 2.16 | 1.17 |
| **Au** | 0.69 | 1.00 | 1.90 | 1.16 | 0.70 | 1.97 | 1.17 | 0.17 | 2.10 | 1.16 |
| **Ni** | 0.65 | 0.28 | 1.78 | 1.17 | 0.56 | 1.92 | 1.16 | 0.12 | 1.74 | 1.18 |
| **Pd** | 0.56 | 0.71 | 1.86 | 1.17 | 0.85 | 1.91 | 1.16 | 0.10 | 1.87 | 1.17 |
| **Pt** | 0.41 | 0.87 | 1.83 | 1.16 | 1.05 | 1.88 | 1.16 | 0.16 | 1.85 | 1.17 |
| **Rh** | 0.80 | -0.33 | 1.81 | 1.18 | 0.63 | 1.84 | 1.17 | 0.21 | 1.84 | 1.18 |
| **Ir** | 0.57 | -0.09 | 1.81 | 1.19 | 0.89 | 1.84 | 1.18 | 0.25 | 1.85 | 1.18 |
| **Ir(5-fold)** | 0.04 | 0.25 | 1.84 | 1.17 | --- | --- | --- | --- | --- | --- |

**Table S4: Structural parameters for the metal-CO systems determined by PBE+U calculations.** Table S4 summarizes the structural parameters (in Å) for these configurations, including the height of the adatom above the average position of the surface Fe-O plane prior to CO adsorption ($z$), the change in the height of the metal atom induced by CO adsorption ($\Delta z$), the Me$_1$-C bond lengths and the inter-molecular C−O bond lengths. The group 11 and 10 metals are lifted from the surface by CO adsorption, whereas Rh and Ir relax towards the surface.

**Computational Results for Group 11 (Cu, Ag, Au)**

As the d-bands of the group 11 elements are completely filled, donation from the 5σ into the d band cannot occur, and the dominant bonding mechanism for CO to the pure metal surfaces is exclusively back-donation from the d-states of the metal into the 2π* orbitals of the molecule. Consequently, the d-COM position is extremely important to the CO binding energy, because this controls how much back-donation will occur.

In Fig. S10, we plot the calculated $E_{ads}(CO)$ as a function of the d-COM for the noble metals. The Bader charge of the metal atom is given as a number next to each data point. Beginning with Ag, we see that the metal surfaces bind CO extremely weakly (< 0.5 eV). This is because the d-COM is very far (4 eV) from the Fermi level, which inhibits back donation. When the metal atom is placed in a 2-fold coordinated adatom site on $Fe_3O_4(001)$, its Bader charge increases to +0.6e and the d-COM moves towards the Fermi level. The Bader charge of +0.6e suggests a +1 oxidation state, which is reasonable because Ag is also 2-fold coordinated to oxygen in $Ag_2O$, where it takes a +1 oxidation state. Since the s electron is given away in the $Ag^{1+}$ state, the d-shell remains completely filled in the 2-fold coordinated geometry and the d-COM position primarily defines the CO adsorption energy. Since it is closer to $E_F$, CO adsorption is stronger than at an Ag metal surface. The stronger bond is also seen in the shorter Ag−C bond length for the $Ag_1/Fe_3O_4(001)$ system (2.00 Å) compared to the Ag(111) surface (2.16 Å) (see Table S3).

If we now look at the $Ag_2O(111)$ surface, we see that two different cation sites are possible; a 2-fold coordinated site and a 1-fold coordinated site. Both have a d-COM close to that of the $Ag_1$-$Fe_3O_4(001)$ system (≈ −3 eV), and the 2-fold site also has a similar Bader charge and CO adsorption energy. The 1-fold site is only partially oxidized, yet exhibits a stronger CO binding energy. This is because the favored linear bonding geometry can be easily attained at such a site, and the adsorption energy is the optimum for the given d-COM position.

In Fig. S11, we show how the structure of the 2-fold $Ag_1/Fe_3O_4(001)$ site and the 1- and 2-fold coordinated sites on $Ag_2O(111)$ are affected by CO adsorption. Darker shades represent the location of the atoms with CO adsorbed, and lighter shades show the positions prior to CO adsorption. The simplest case to understand is the 1-fold site. Here, the Ag atom is undercoordinated with respect to the bulk oxide, and thus offers a free ligand site to the CO, as strong as it can be for the particular d-COM position. In contrast, both of the 2-fold sites are already close to the preferred 2-fold linear coordination (with oxygen), and the CO molecule must compete with the oxide to bind the Ag atom. The $Ag_1/Fe_3O_4(001)$ SAC site is located centrally between the two lattice oxygen atoms, and is lifted away from the surface by 0.29 Å when CO adsorbs. The Ag−O bonds are elongated, but remain equal in length. The structural distortion costs energy, and the system comes to equilibrium when the energetic gain from binding the CO molecule balances the cost of weakening the Ag-$O_{surface}$ bonds. At the 2-fold site on $Ag_2O(111)$ the situation is somewhat different because the oxygen atoms to which the Ag is bound are inequivalent. When CO adsorbs, the lower-coordinated of the surface oxygen atoms relaxes strongly, while the higher-coordinated oxygen remains essentially in place. Since elongating the former Ag−O bond is relatively inexpensive the Ag is lifted strongly away from the surface by 1.23 Å. Ultimately, the energetic cost of the distortions is similar, and the 2-fold coordinated Ag sites exhibit a similar CO adsorption energy.

Turning now to Cu, we see that the d-COM for the metal surfaces are significantly closer to $E_F$ than they were for Ag, which enhances back donation, and explains the stronger CO binding. Placing the Cu atom in 2-fold coordination to the metal oxide surface again leads to a 1+-like state, and a d-COM shifted towards $E_F$. However, in this case, it does not translate into a stronger CO adsorption due to structural distortions in the system: The stronger Cu-CO bond at the $Cu_1/Fe_3O_4(001)$ site can be seen in the shorter Cu−C bond-length (1.81 Å) compared to the Cu(111) surface (1.86 Å). Nevertheless, the CO adsorption energy is weaker than that expected on the basis of electronic structure because the $Cu_1$ adatom is lifted by 0.53 Å from the surface upon CO adsorption. This strong CO-induced distortion of the system reduces the total energy. For $Cu_1/Fe_3O_4(001)$, the increased Cu−CO bond strength arising from the electronic structure is almost exactly compensated by the energy lost by the structural distortion. Thus, Cu appears to behave like the metal surface in TPD, but this is a coincidence.

Finally, we consider Au. Again, CO adsorbs weakly at the metal surfaces because back-donation dominates and the d-COM is far from the Fermi level. When an Au metal atom is placed at the $Fe_3O_4(001)$ surface, the d-COM shifts towards $E_F$ and the CO binding energy increases, as seen for Ag and Cu. The $Au_2O(111)$ reference was calculated for comparison, but we note that it is marginally stable with our computational setup, and is known to be unstable in experiment. Thus, we do not draw strong conclusions from these data points.

It is interesting that the extent of the CO-induced lifting differs dramatically depending on the system. We attribute this to the different CO and O affinity of the metals in question. For a qualitative picture, we calculated the binding energy for CO and O (i.e. ½ $O_2$) to the gas-phase metal atoms (see Fig. S12). Such gas phase calculations do not perfectly represent the charged adatoms, but as will be shown in the subsequent sections, this plot can help to understand the extent of CO-induced lifting, as well as the CO-induced sintering that occurs for Pd and Pt (*23, 26*).

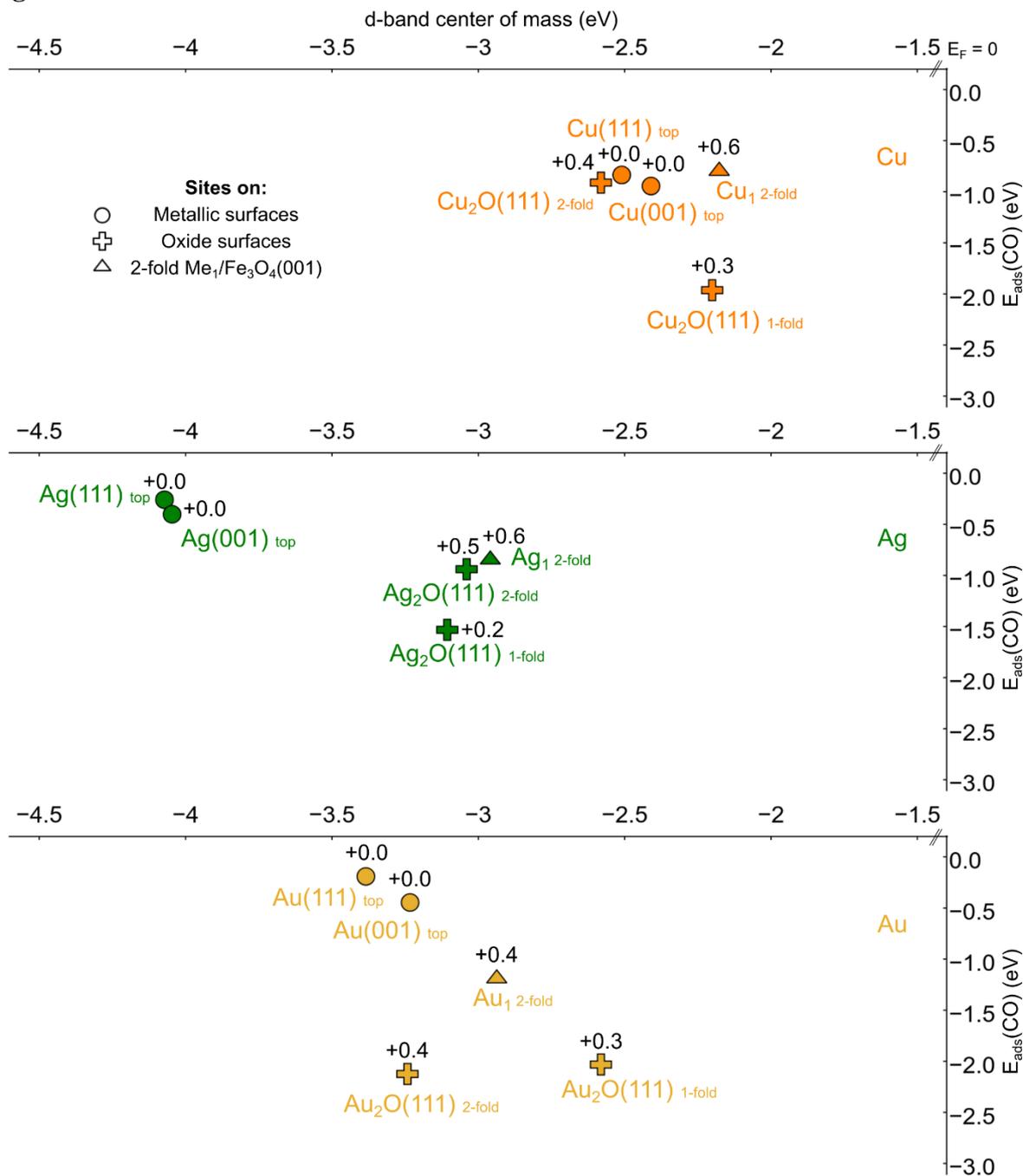

**Figure S10: $E_{ads}$(CO) as a function of d-band center of mass for group 11 (Cu, Ag and Au).** Data are included for metal surfaces (circles), oxide surface (crosses) and 2-fold SAC sites on $Fe_3O_4(001)$ (triangles). Bader charges are given in black text next to each data point. For reference, the Bader charges of bulk cations in the metal oxides are: $Cu_2O$: 0.52 ($Cu^{1+}$), $Ag_2O$: 0.45 ($Ag^{1+}$) and $Au_2O$: 0.42 ($Au^{1+}$).

**Figure S11**

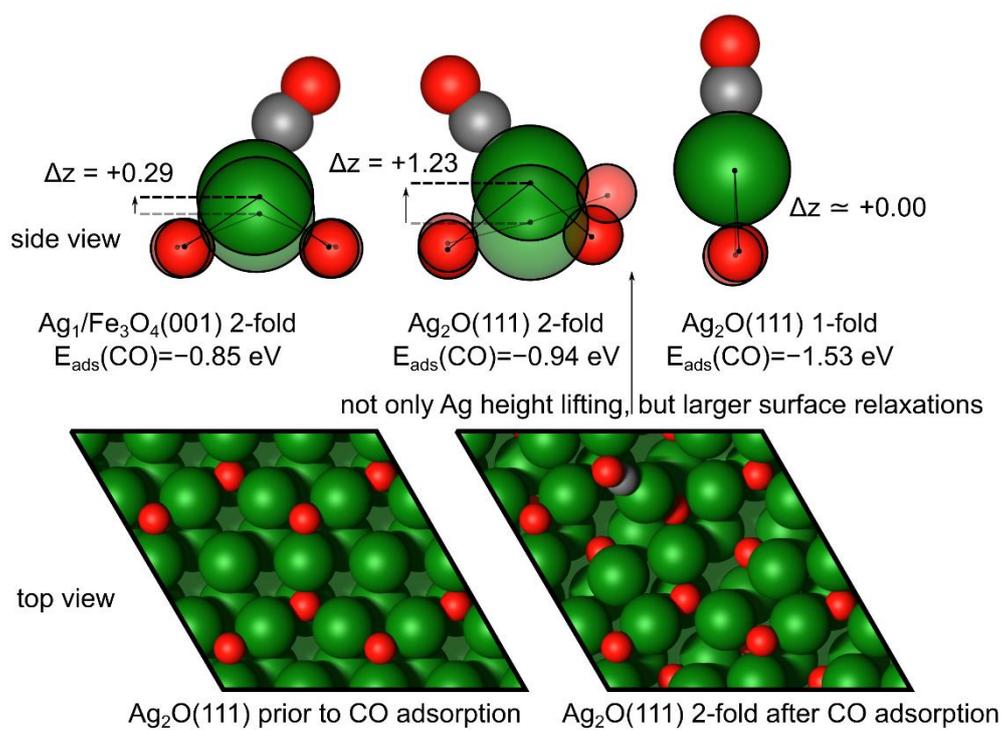

Δz = +0.29

Δz = +1.23

side view

Δz ≃ +0.00

$Ag_1/Fe_3O_4(001)$ 2-fold
$E_{ads}(CO)=-0.85$ eV

$Ag_2O(111)$ 2-fold
$E_{ads}(CO)=-0.94$ eV

$Ag_2O(111)$ 1-fold
$E_{ads}(CO)=-1.53$ eV

not only Ag height lifting, but larger surface relaxations

top view

$Ag_2O(111)$ prior to CO adsorption

$Ag_2O(111)$ 2-fold after CO adsorption

**Figure S11:** Relaxations occurring at the 2-fold $Ag_1/Fe_3O_4(001)$ site in comparison with the 1- and 2- fold sites on $Ag_2O(111)$.

**Figure S12**

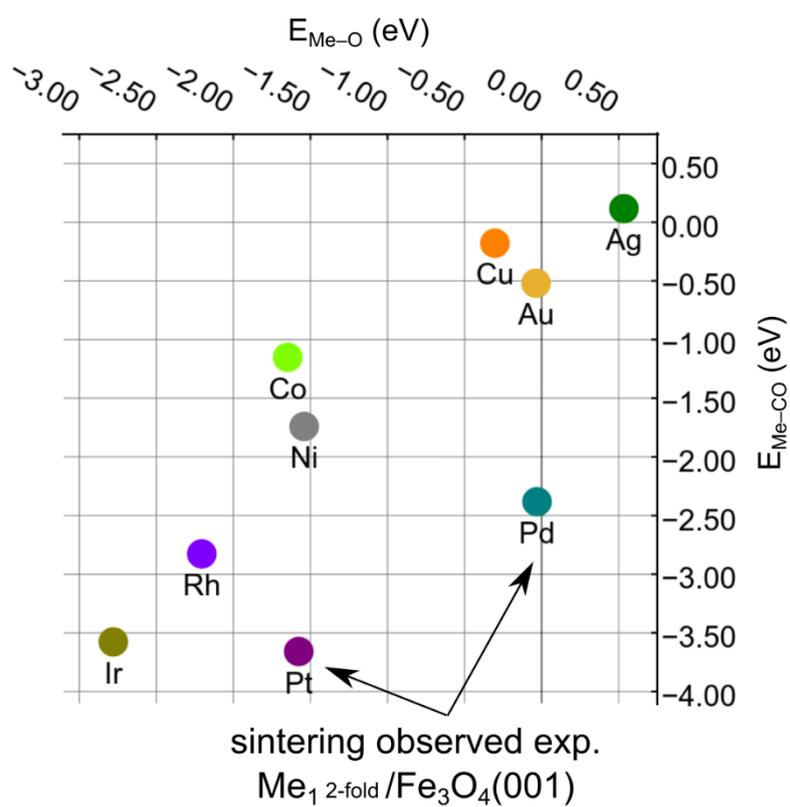

**Figure S12:** Binding energies for O and CO on a Me atom in the gas phase, Me = [Cu, Ag, Au, Ni, Pd, Pt, Co, Rh, Ir] (energies of ½ $E_{O2}$ and $E_{CO}$ used as references).

**Computational Results for Group 10 (Ni, Pd, Pt)**

In group 10, the outermost d-shell is not completely full in the metallic state, so 5σ donation into the metal surface can also contribute to the CO binding energy (*41*) (see Fig. S9). Figure S13 shows the calculated $E_{ads}$(CO) as a function of the d-COM for the group 10 metals. For the Ni metal surfaces, the d-COM is relatively close to $E_F$ ($\approx$ −1.6 eV), and since d-2π* back donation still dominates, CO binding is strong. When the Ni atom is placed in the 2-fold site on $Fe_3O_4$(001), the atom becomes 1+ (we assign the unusual 1+ state on the basis of the Bader charge). The oxidation of the metal atom reduces the d-band filling from that of the pure metal, and the d-COM shifts away from $E_F$. The CO adsorption energy is then weaker at the 2-fold coordinated adatom site. Experiments have shown that this site is unstable with respect to substitution of a 5-fold coordinated surface Fe atom at room temperature. Calculations for this configuration show that the d-COM shifts still further from $E_F$, resulting in an extremely weak CO adsorption energy. Indeed, the structure, Bader charge, and CO adsorption energy resemble calculations for NiO(100), where the bonding has been shown to be mainly electrostatic (*43*). This shows that metal adatoms can be expected to behave like the stable oxide when the geometry is similar, but that metastable geometries such as the 2-fold coordinated site studied here can yield unique properties.

The theoretical details for Pd and Pt are discussed next, but it is important to note that $Pt_1$ and $Pd_1$ adatoms on $Fe_3O_4$(001) (*23, 26*) have been observed to sinter rapidly upon exposure to CO in experiment. It can be seen in Fig. S12 that Pt and Pd have a stronger affinity for CO compared to O than other elements, and this leads to large increases in the height of the adatom when CO adsorbs ($\Delta z_{Pt}$ = 0.87 Å, $\Delta z_{Pd}$ = 0.71 Å). The weakening of the metal-support interaction lowers the diffusion barrier, which enables rapid agglomeration into nanoparticles at room temperature. In spite of this, it is instructive to consider the behavior of the adatoms in comparison to the metal and metal oxide surfaces.

Pt is an example of a metal where the 5σ-d donation contribution becomes comparable in importance to back-donation (*40*). As we move away from the $E_F$, the back-donation is again weakened, but this is compensated by a strengthening of the donation mechanism. This improvement of the donation has two causes: the 5σ-d split is reduced facilitating the corresponding hybridization, and the higher oxidation reduces the d-filling, allowing more electron density to be donated from the 5σ orbital of the CO molecule. In consequence, we obtain a roughly constant CO adsorption energy irrespective of the d-COM position and Bader charge. The PtO(100) surface provides an exception to this behavior because the Pt cation already exhibits the favored bulk coordination, and thus offers no free ligand site to bind the CO molecule. In this situation the CO binding energy is extremely weak. We note that this geometry is similar to that observed for $Pt^{2+}$ occupying step edge sites on $CeO_2$(111) (*67*), which were found to be extremely stable against thermal sintering, but catalytically inert.

The data for Pd are interesting because the 2-fold $Me_1$/$Fe_3O_4$(001) systems appears electronically similar to the metal surfaces, despite the 1+ oxidation state. As was seen for Ni, the d-COM at the 5-fold $Pd_1$/$Fe_3O_4$(100) site exhibits a strong shift away from $E_F$, close to that seen for the PdO(100) site. PdO(100) is similar to PtO(100), in that the Pd atom is coordinatively saturated. As a consequence, the CO binding energy is extremely weak.

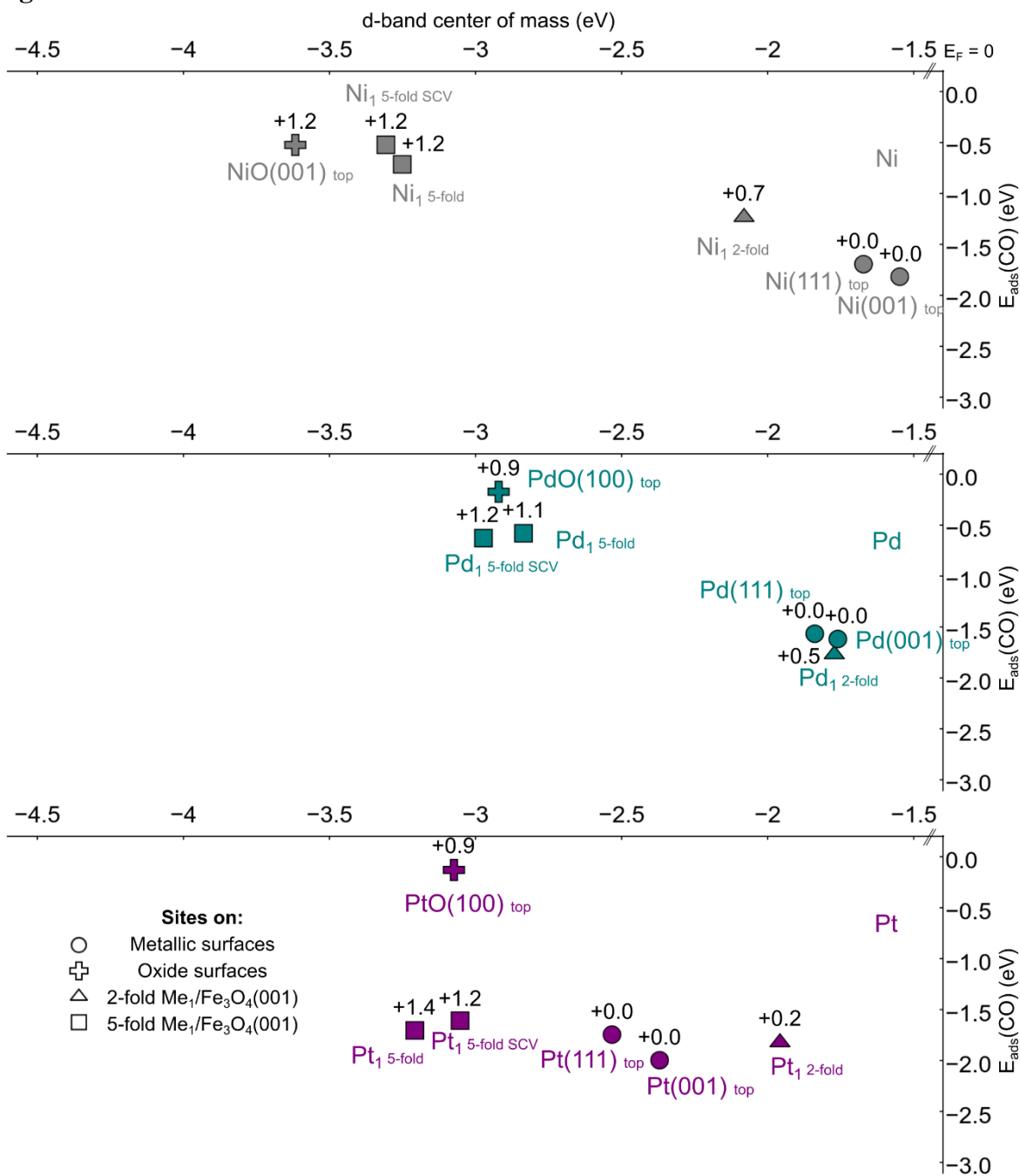

**Figure S13: E_ads(CO) as a function of d-band center of mass for group 10 (Ni, Pd and Pt)**. Adsorption energies were calculated for sites on metal surfaces (circles), oxide surface (crosses) and SAC on $Fe_3O_4$(001), 2-fold (triangles) and 5-fold (squares) sites. Bader charges are shown next to the data points, and can be compared to bulk references for: NiO: 1.18 ($Ni^{2+}$), PdO: 0.83 ($Pd^{2+}$), PtO: 0.83 ($Pt^{2+}$).

**Computational Results for Group 9 (Rh, Ir)**

In Fig. S14, we plot the calculated $E_{ads}(CO)$ as a function of the d-COM for the group 9 metals Rh and Ir. Cobalt is not included because the metal 2-fold $Co_1/Fe_3O_4(001)$ adatoms are not stable, and have been observed to incorporate into the $Fe_3O_4(001)$ support already at room temperature (*25*).

CO adsorption at the 2-fold $Rh_1/Fe_3O_4(001)$ and $Ir_1/Fe_3O_4(001)$ systems is found to be slightly stronger than at the respective metal surfaces, even though the d-COM positions are similar. We have previously shown that the metal atom relaxes towards the surface and forms a bond to a subsurface oxygen atom when CO adsorbs (*29*), allowing the metal atom to achieve a pseudo-square planar configuration typical of Rh(I) and Ir(I) complexes, improving therefore E(CO). In the TPD experiment, CO desorption actually occurs from a 5-fold coordinated site for Ir (*29*), because the metal incorporates into the surface before the CO molecule desorbs. In the case of Rh, CO on the 5-fold coordinated SAC sites is slightly less strongly bound than on the 2-fold $Rh_1/Fe_3O_4(001)$. For Ir, CO on the 5-fold coordinated SAC site is slightly stronger than the on 2-fold $Ir_1/Fe_3O_4(001)$ site. This likely reflects the greater importance of the donation interaction for Ir, because the d-COM position is further from $E_F$. In both cases, the 5-fold coordinated sites exhibit a similar d-COM, Bader charge, and CO adsorption energy as 5-fold coordinated sites at the corresponding metal oxide surfaces.

**Figure S14**

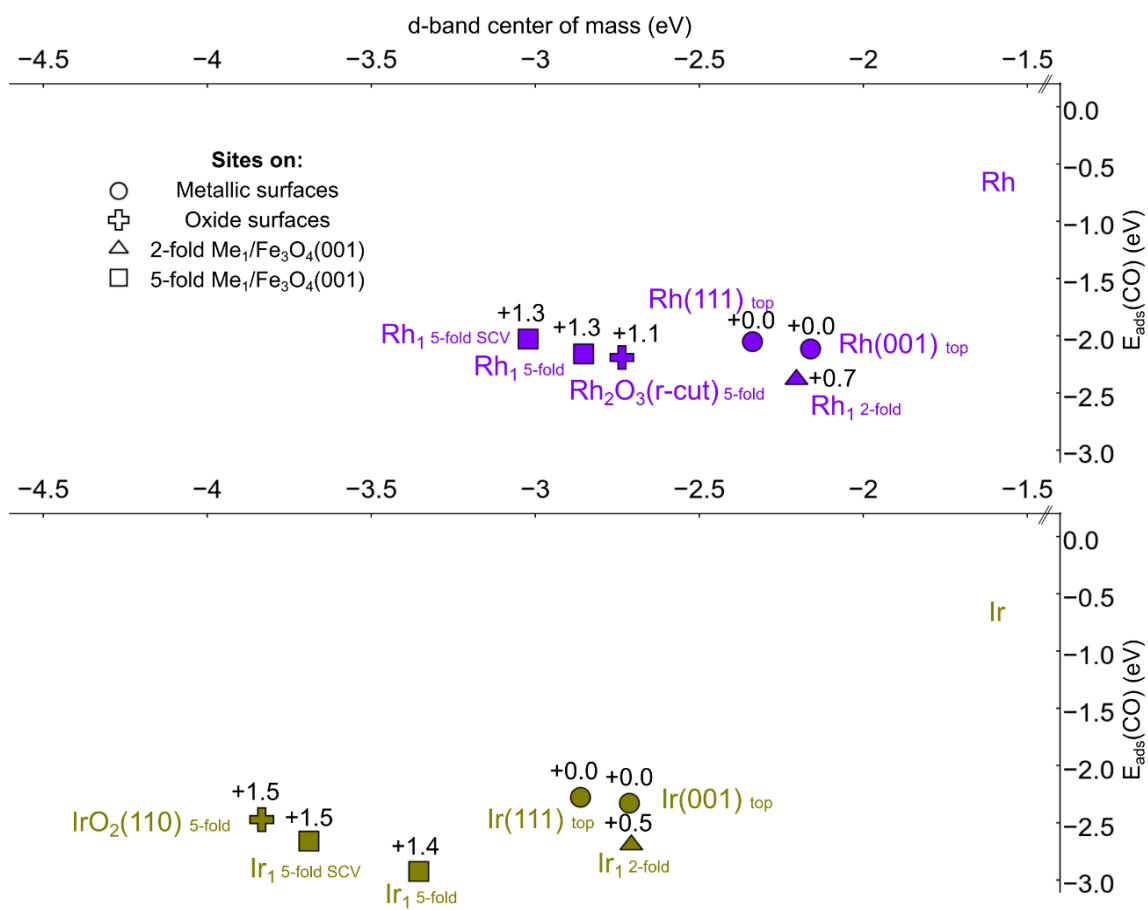

**Figure S14: $E_{ads}(CO)$ as a function of d-band center of mass for group 9 (Rh and Ir).** Calculated for sites on metal surfaces (circles), oxide surface (crosses) and SAC on $Fe_3O_4(001)$, 2-fold (triangles) and 5-fold (squares) sites, when available. Bader charges are given for each site. Bulk metal oxide Bader charges as a benchmark: $RhO_2$: 1.52 ($Rh^{4+}$), $IrO_2$: 1.65 ($Ir^{4+}$).

**Table S5**

| Core level | Experiment (Fig S2) | DFT |
|:---:|:---:|:---:|
| Cu2p | +0.80 | +0.61 |
| Ni2p | +0.80 | +0.83 |
| Rh3d | +1.00 | +1.34 |

**Table S5: CO induced core level shift (in eV) compared between theory and experiment.** The relative shift of the metal 2p (for Ni and Cu) or 3d (Rh) levels are calculated upon CO adsorption. A full electron was removed from the respective core and placed in the valence band. The hole is screened by allowing electronic relaxations, leading to the so-called final state approximation. Note, however, that since we use VASP with a pseudo-potential to simulate core electrons, results concerning core level shifts have to be taken with care. Overall, the relative trends are well reproduced, considering the level of approximations and setup used.

# Table S6

| Metal | Surface | $E_{des}$ (kJ/mol) | $E_{des}$ (eV) | Method | Prefactor | Fractional. coverage | CO site (assumed) | Metal coord. | Ref. |
|---|---|---|---|---|---|---|---|---|---|
| Cu | 111 | 47 | **0.49** | TPD | 1E+13 | <0.3 | on top | 9 | (68) |
| Cu | 100 | 51 | **0.53** | TPD | 1E+13 | <0.3 | on top | 8 | (68) |
| Cu | 110 | 52 | **0.54** | TPD | 1E+13 | 0.1 | | 6 | (69) |
| Cu | 110 | 54 | **0.56** | TPD | 1E+13 | <0.3 | on top | 6 | (68) |
| Cu | 211 | 58 | **0.60** | TPD | 1E+13 | <0.3 | on top | 7 | (68) |
| Cu | 221 | 58 | **0.60** | TPD | 1E+13 | <0.3 | on top | 7 | (68) |
| Cu | 532 | 57 | **0.59** | TPD | 1E+13 | <0.3 | on top | 6 | (68) |
| Cu | sputtered 100 | 59 | **0.61** | TPD | 1E+13 | <0.3 | likely on top | 6-7 | (68) |
| Cu | polycrystalline | 58 | **0.60** | TPD | 1E+13 | <0.3 | likely on top | 6-7 | (68) |
| Ag | 111 | 27 | **0.28** | isotherms | | | | 9 | (70) |
| Au | 111 | 17 | **0.18** | TPD | 2E+15 | | | 9 | (71) |
| Au | 110 | 38 | **0.39** | TPD | 3E+13 | 0.12 | | | (72) |
| Au | 211 | 50 | **0.52** | TPD | 1E+13 | zero cov. limit | | | (73) |
| Au | 332 | 48 | **0.50** | TPD | 1E+13 | <0.2 | | | (74) |
| Ni | 111 | 126 | **1.31** | TPD | 1E+15 | 0.1 | bridge | 9 | (75) |
| Ni | 111 | 130 | **1.34** | TPD | 2E+14 | 0.05 | | 9 | (76) |
| Ni | 110 | 106 | **1.10** | TPD | 1E+13 | <0.7 | bridge | | (77) |
| Ni | 110 | 138 | **1.43** | TPD | 8.5E+15 | 0.03 | | | (78) |
| Ni | 110 | 116 | **1.2** | TPD | 1E+13 | | | | (79) |
| Pd | 110 | 132 | **1.37** | TPD | 5E+13 | 0.1 | | | (80) |
| Pd | 100 | 154 | **1.60** | TPD | 2E+16 | 0.15 | bridge | 8 | (81) |
| Pd | 112 | 143 | **1.48** | TPD | 1E+15 | zero cov. limit | | | (82) |
| Pd | 111 | 149 | **1.54** | TPD | 1e13.5 | zero cov. limit | 3 fold hollow | 9 | (83) |
| Pt | 110 | 129 | **1.34** | TPD | 1E+13 | zero cov. limit | 4-fold | 6 | (84) |
| Pt | 110 | 147 | **1.52** | TPD | 6E+14 | zero cov. limit | | | (85) |
| Pt | 111 | 138 | **1.43** | TPD | 1E+13 | zero cov. limit | | 9 | (86) |
| Pt | 111 | 139 | **1.44** | TPD | 4E+15 | zero cov. limit | | 9 | (87) |
| Pt | 111 | 145 | **1.50** | isotherms | | | | 9 | (88) |
| Pt | 111 | 134 | **1.39** | laser induced desorption | 1E+14 | zero cov. limit | | | (88) |
| Pt | 335 (edge) | 134 | **1.39** | TPD | 7E+12 | 0.015 | edge site | | (90) |
| Pt | 335 (terrace) | 113 | **1.17** | TPD | 3E+12 | 0.015 | terrace site | | (90) |
| Rh | 110 | 132 | **1.37** | TPD | 1E+13 | zero cov. limit | | | (91) |
| Rh | 111 | 134 | **1.39** | TPD | 1E+13 | zero cov. limit | | 9 | (92) |
| Rh | polycrystalline | 134 | **1.39** | TPD | 1E+13 | dose 0.08 L | | | (93) |
| Ir | 110 | 146 | **1.51** | TPD | 1E+13 | zero cov. limit | | | (94) |
| Ir | 111 | 134 | **1.39** | TPD | 1E+13 | 0.1 | | 9 | (95) |

**Table S6:** Summary of experimental CO binding energies on different metal surfaces. Grey shaded data were used in the construction of Figure 2.

**Table S7**

| Metal | ΔE(eV) (going from twofold to fivefold site) |
|---|---|
| Cu | 0.65 |
| Ag | 1.14 |
| Au | 0.84 |
| Ni | -0.04 |
| Pd | -0.23 |
| Pt | 0.02 |
| Rh | -1.05 |
| Ir | -0.97 |

**Table S7**: Calculation of the energy gained in moving a twofold coordinated adatom to a fivefold coordinated "incorporated" site substituting surface Fe within the $Fe_3O_4(001)$ lattice. Note that negative is an energy gain.